\documentclass[onecolumn,groupedaddress,showkeys]{article}
\usepackage{float}
\usepackage{graphicx}
\usepackage{amssymb}
\usepackage{authblk}
\usepackage{amsmath}
\usepackage{multirow}
\usepackage{hhline}
\usepackage{caption}
\usepackage{subcaption}
\usepackage{xcolor}
\usepackage{gensymb}
\usepackage[breaklinks]{hyperref}
\hypersetup{colorlinks,urlcolor=black,citecolor=black,linkcolor=black,filecolor=black}
\usepackage{anysize}
\usepackage[left=1cm,right=1cm,top=1cm,bottom=1cm,includehead,includefoot]{geometry}
\newcommand{\pref}[1]{(\ref{#1})}

\date{September 2023}

\begin{document}
\title{Studying ECG signals using nonlinear oscillators and Genetic Algorithm}
\author{Sourav Chowdhury\thanks{email: chowdhury95sourav@gmail.com}\qquad Apratim Ghosal\thanks{email: apratim99.ag@gmail.com}\qquad Suparna Roychowdhury\thanks{email: suparna@sxccal.edu}\qquad Indranath Chaudhuri\thanks{email: indranath@sxccal.edu}}
	\affil{Department of Physics, St. Xavier's College (Autonomous)\\
		30 Mother Teresa Sarani, Kolkata-700016, West Bengal, India}

\maketitle

\begin{abstract}
\noindent  Cardiovascular diseases are the leading cause of death and disability in the world and thus their detection is extremely important as early as possible so that it can be prognosed and managed appropriately. Hence, electrophysiological models dealing with cardiac conduction are critically important in the field of interdisciplinary sciences. The primary aim of this paper is to reproduce a normal sinus rhythm ECG waveform which will act as the baseline for fitting and then fit any clinical ECG waveform that does not deviate much from normal sinus rhythm. To reproduce the ECG, we modeled the pacemaker complex using three coupled van der Pol (VDP) oscillators with appropriate delays to generate the action potentials. These action potentials are responsible for the excitation of the non-pacemaker cells of the atria and ventricles whose electrical activity gets recorded as the ECG signal. The ECG signal is composed of a periodic set of individual waves corresponding to atrial and ventricular contraction and relaxation. These waves are modeled with the help of four FitzHugh-Nagumo (FHN) equations with impulses corresponding to the action potentials generated by the pacemaker cells. After the successful reproduction of a normal sinus rhythm ECG, we have developed a framework where we have used genetic algorithm (GA) to fit a given clinical ECG data with parameters belonging to the above mentioned system of delay differential equations (DDEs). The GA framework has enabled us to fit ECG data representing different cardiac conditions reasonably well. We aim to use this work to get a better understanding of the cardiac conduction system and cardiovascular diseases which will help humanity in the future.
\end{abstract}


\section{Introduction}
Cardiovascular disease (CVD) is one of the major reasons for death and disability in 
the world \cite{vaduganathan2022global,tsao2023heart}. An estimated 20.5 million deaths were caused by CVDs which constitutes about a third of the global annual death in 2021 \cite{pineiro2023world}.  According to the World Health Organisation, India constitutes about one-fifth of these global deaths and an alarming number of such cases have been observed in the younger population  \cite{kalra2023burgeoning,kumar2020cardiovascular}. The reason for such high rates among Indians can be attributed to a variety of factors ranging from diseases like diabetes, hypertension, low-activity sedentary lifestyle, psychological stress, smoking, pollution and the like, but it is currently unknown as to why the risk of CVDs is increasing in the younger population. To understand these reasons, the need of models explaining the cardiovascular conduction along with electrophysiological activity has become of prime importance in the field of medicine and interdisciplinary sciences. \\

\noindent These models which connect cardiovascular conduction along with electrophysiological activity has originated from existing studies in skeletal muscles. This is due to the fact that there exists a close link between cardiac pacemaker cells and the sarcomeres which are fundamental units producing muscle contraction \cite{niederer2019short}. The first tissue-based model was developed by A.V. Hill \cite{hill1938heat} in 1938. In his work, he discussed `a rheological model relating the release of energy by the muscle, the mechanical work performed by the muscle and the force of contraction and the velocity of shortening' \cite{hill1938heat} which was experimentally verified by Huxley and Simmons in 1957 \cite{huxley1957muscle}. Subsequently, biochemical modeling of contraction and muscle activation was also performed by various others linking Hill and Huxley-Simmons' observations \cite{julian1969activation,saeki1980transient}. Following these studies, some early cardiac models were also developed \cite{parmley1967series,fung1970mathematical}. In terms of modeling the whole heart activity, Arts \textit{et al.,} \cite{arts1979model} assumed the left ventricle to be a thick-walled concentric cylindrical shell and the myocardium to be anisotropic. They simulated a model that was able to find the pressure-volume relationship of the left ventricle and another relationship between the transmural distribution of sarcomere length and fiber stress. This model was further refined by Guccione \textit{et al.,} \cite{guccione1991passive,guccione1995finite} who introduced `a finite element method to solve the deformation mechanics equations in a circumferentially symmetric left ventricle'. Next, he successfully predicted the stress and verified their findings by using experimental data \cite{aguado2011patient,sermesant2012patient,crozier2016relative,kayvanpour2015towards}.   \\

\noindent Electrophysiological models involve the electrical activity of the myocardium and the transmembrane ion currents which are responsible for the electrical activity of the cardiac muscles. One class of model is usually concerned with the functioning of the heart on a cellular or tissue level. They involve modeling each action potential generating component of the heart separately. These models take into account all of the ion diffusion currents and ion concentrations both inside and outside the cardiac cell. The diffusion currents are based on the Hodgkin-Huxley equations \cite{hodgkin1952quantitative} and are classed as reaction-diffusion type models. Since a large number of such equations need to be solved, they usually involve a heavy usage of computational tools. The first true whole organ model was developed by Nash and Hunter \cite{nash2000computational} which combined ventricular contraction and fiber orientation. O'Hara \textit{et al.,} \cite{o2011simulation} were successful in simulating the healthy human cardiac ventricular action potential and also verified it experimentally. Several tissue-level works have also been successfully developed such as Ten Tusscher \textit{et al.,} \cite{ten2008modelling} which uses a PDE-based monodomain model \cite{keener2009mathematical}.  A huge number of such models are available which have variables ranging from 4-variable Bueno-Orovio-Cherry-Fenton ventricular model \cite{bueno2008minimal} to 67-variable Iyer-Mazhari-Winslow ventricular model \cite{iyer2004computational}. The single-cell models when integrated with machine learning have been successful in predicting drug-induced arrhythmia \cite{lancaster2016improved} and are gaining traction in recent years for testing new drugs seeking regulatory approval. Tissue-based models have also been used in the improvement of stem cell therapy in heart muscles \cite{mayourian2017experimental}. Nowadays Markov chain models and state diagrams are being widely used for more accurately simulating complex ion channel gating behaviors \cite{doerschuk1990modelling,smith2004development}. Lumens \textit{et al.,} \cite{lumens2015differentiating} has developed a series of models that is widely used to study clinical data using a 1D and 2D representation of the ventricles. Following these models, some patient-specific models have also been developed in recent years to mainly aid in cardiac resynchronization therapy (CRTs) which is used to regulate proper rhythm of the heart using an external pacemaker \cite{sermesant2012patient,crozier2016relative}. Recently cardiac imaging studies use artificial intelligence for beat-to-beat assessment \cite{ouyang2020video} and early detection of cardiac disorders \cite{jiwani2021novel,zhang2020real}. \\

\noindent
The other class of electrophysiological models treat the action potentials generated by each pacemaker and non-pacemaker cells grouped together separately as a single unit. They try to generate a synthetic ECG waveform without going into the detailed internal structure of the heart. These class of models are generally ODE-based where the heart is compartmentalized into pacemaker and non-pacemaker cardiac cell blocks. The maximum emphasis is given on the generation of the specific action potentials and how they couple with each other, thereby producing a synchronized response of the atria and ventricles which is obtained as the final ECG waveform. Each of these compartments are modeled by coupled modified van der Pol oscillators pioneered by Balthasar van der Pol and van der Mark \cite{van1928lxxii}.  Ryzhii \textit{et al.,} \cite{ryzhii2014heterogeneous}, whose work is the primary inspiration behind this paper have developed an ODE-based model with delays (DDE). They model the pacemaker cell compartments using coupled modified van der Pol oscillators and the non-pacemaker cells of the atria and ventricles which behave as excitable media using the FitzHugh-Nagumo (FHN) equations \cite{fitzhugh1961impulses,nagumo1962active}. The FHN equations are nothing but a 2D analogue of the Hodgkin-Huxley equations \cite{hodgkin1952quantitative}. Their work was a refinement of the earlier works done by di Bernardo \textit{et al.,} \cite{di1998model}, Sato \textit{et al.,} \cite{sato1994bonhoeffer}, Grudzi{\'n}ski \textit{et al.,} \cite{grudzinski2004modeling}, Gois \textit{et al.,} \cite{gois2009analysis} and many others. 
\\


\noindent
Most of these papers have been successful in recreating the ECG signal but none of them have tried to properly fit the model with raw ECG data. In this paper, we have used the Ryzhii model and its parameters to generate a synthetic ECG signal replicating a normal heart rate. We have also successfully modified the Ryzhii model with a suitable delay term (henceforth our model will be referred to as $\tau_{\scriptscriptstyle T}$-Ryzhii Model) to find an optimized solution which fits different kinds of cardiac ECG signals reasonably well. For optimizing the parameters of the $\tau_{\scriptscriptstyle T}$-Ryzhii model, we have further developed a framework with genetic algorithm (GA). This produces a good fit to the data representing normal sinus rhythm and datasets representing extreme ends of the normal sinus rhythm, i.e., sinus tachycardia and sinus bradycardia.  Finally, we have also fitted datasets exhibiting pathological conditions which do not deviate much from normal sinus-type waveforms with our improved $\tau_{\scriptscriptstyle T}$-Ryzhii model. \\

\noindent The paper has been organized as follows: In Sec. \pref{Cardiac Conduction System} we have defined our system and various technical terms pertaining to it. In Sec. \pref{Formulation} we have discussed the Ryzhii model in detail and the reasons for modifying the existing model. In Sec. \pref{Mod_fit} we have a detailed discussion of the development of the genetic algorithm (GA) which is used to fit the ECG waveform data. In Sec. \pref{results} we have displayed the ECG waveform fits and our findings regarding each fit. Finally, we have listed the concluding remarks in Sec. \pref{conclusion}.



\section{Cardiac Conduction System}
\label{Cardiac Conduction System}

Here we have discussed the workings of cardiac conduction system which comprises of the pacemaker cell complexes and the non-pacemaker cells of the atria and the ventricles of the heart, the action potentials which are generated by these cells and the formation of the ECG signal.

\subsection{Pacemakers: Origin of Action Potentials }
Each heartbeat originates as an electrical impulse from a group of specialized cells in the heart called pacemaker cells (refer to Fig.~(\ref{fig:ecg_gen}) for a detailed illustration). These cells have the unique ability of producing electrical impulses by depolarizing spontaneously and forwarding them via gap junctions of the adjacent cardiac myocytes (which include both pacemaker as well as non-pacemaker cells of the atria and ventricles). These impulses are called action potentials (AP), the rate of which is controlled by the autonomic nervous system (ANS) \cite{purves2019neurosciences}.  The rate of firing or frequency of these action potentials also control the heart rate. The highest concentration of these pacemaker cells are found in the Sinoatrial (SA) node, and therefore it is called the primary pacemaker of the heart. The intrinsic frequency of depolarization  of the SA node is between 60-100 beats per minute (bpm) \cite{litfl,kusumoto2020ecg,wagner2001marriott}. The impulses from the SA node then reach the atrioventricular (AV) node which works as the secondary pacemaker and has an intrinsic depolarization rate of 40-60 bpm \cite{litfl,kusumoto2020ecg,wagner2001marriott}. There is also the Bundle of His, whose left and right bundle branches and the Purkinje fibers (HP) has intrinsic depolarization rate of 30-40 bpm \cite{litfl,kusumoto2020ecg,wagner2001marriott}. Since the SA node has the fastest rate of depolarization, it sets the heart rate and drives the other secondary pacemakers (Atrioventricular node and the His-Purkinje complex) to depolarize at the rate set by itself. This helps in bypassing the natural frequency of depolarization of the secondary pacemakers and thus synchronizes the entire process at the rate set by the SA node. If one of the higher frequency pacemakers fail to produce action potentials, the secondary pacemakers will dictate the heart rate. The primary aim of this model is to properly represent this synchronized working of the pacemakers. The electrical activity given by the eventual contraction of the contractile myocytes, which are the non-pacemaker cells, superposed on the signal received from the pacemaker cells form the ECG signal. 

\subsection{Action Potential Generation}

An action potential is defined as a change in the potential of the cell membrane of the cardiac myocytes (voltage of cell membrane measured with respect to the interior voltage of the cardiac myocyte) caused due to influx and outflux of ions ($\text{K}^{+}$, $\text{Na}^{+}$, $\text{Ca}^{2+}$) regulated by respective ion channels which are specialized membrane proteins. When the cell is not electrically excited, the interior of the myocyte has a potential of about $-$90mV with respect to the exterior (resting membrane potential). The interior of the cell at rest contains mostly $\text{K}^{+}$ ions while the exterior has high concentration of  $\text{Na}^{+}$ ions which are held in equilibrium by the respective pumps in the cell membrane \cite{santana2010does}. \\

\noindent The action potential generated by the pacemaker cells differ from those which are conducted by the contractile cardiac myocytes (non-pacemaker cells) of the atria and the ventricles in their structure and the generation mechanism. The pacemaker cells produce slow response action potentials while the atrial and ventricular myocytes produce fast response action potential. 

\begin{figure}[H]
    \begin{subfigure}{.5\textwidth}
        \centering
        \includegraphics[scale=0.049]{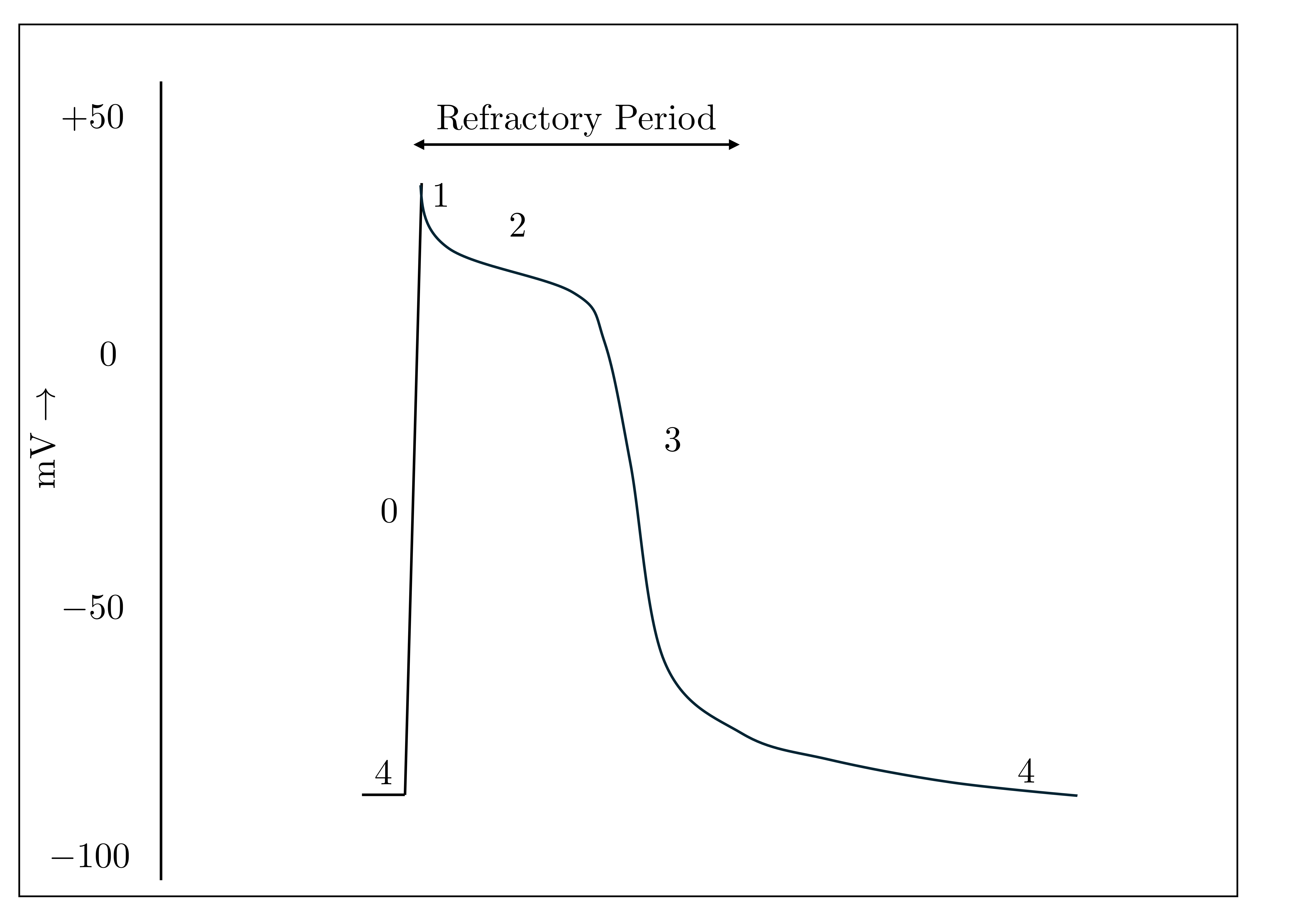}
        \caption{}
    \end{subfigure}
    \begin{subfigure}{.5\textwidth}
        \centering
        \includegraphics[scale=0.056]{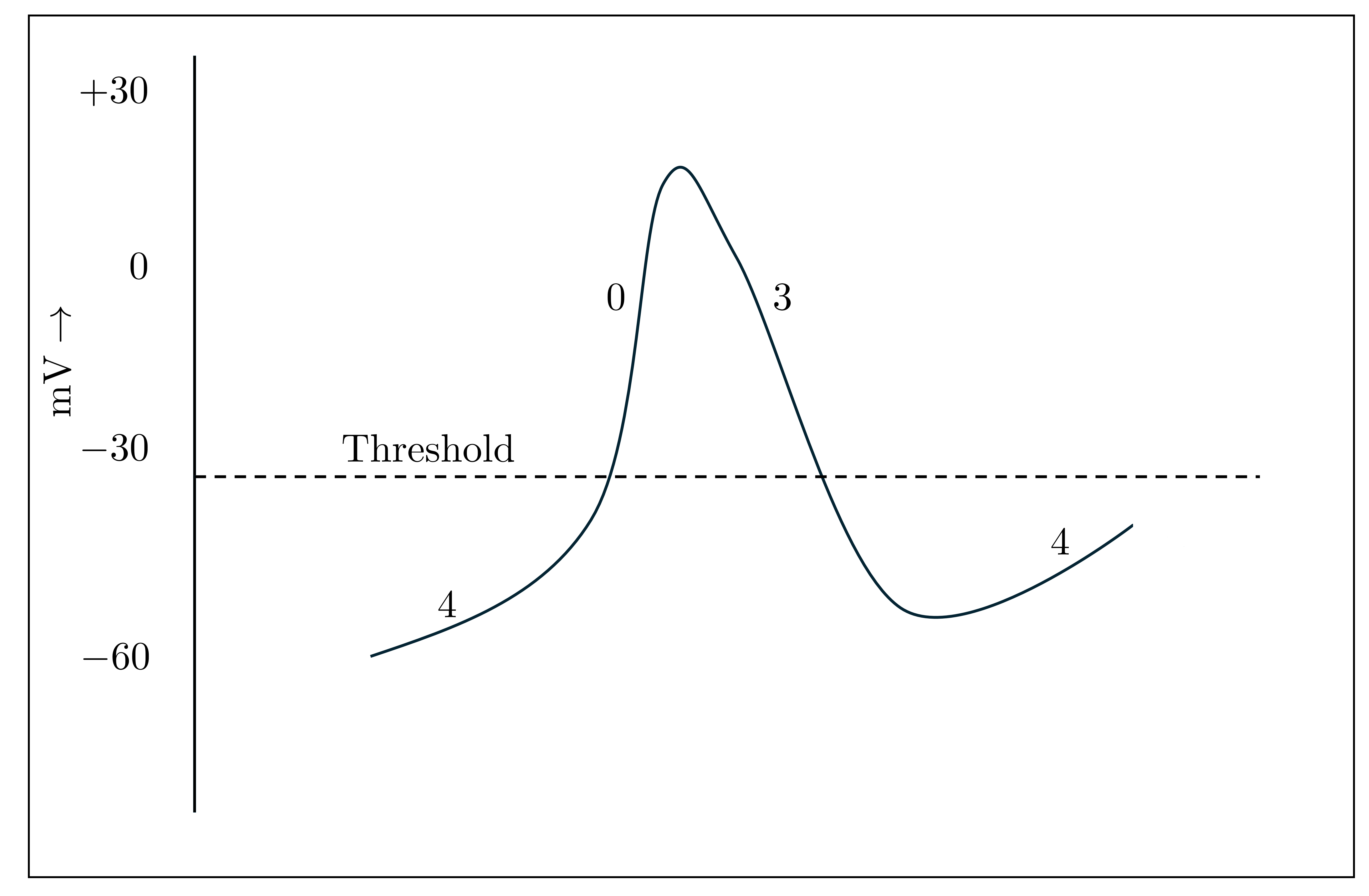}
        \caption{}
    \end{subfigure}
    \caption{Types of Action Potential. (a) Fast Response Action Potential, and (b) Slow Response Action Potential. } 
    \label{fig:AP}
\end{figure}

The action potential is generated in four phases (see Fig.~\pref{fig:AP}): 
\begin{itemize}
    \item Phase 4: The non-pacemaker cell membrane is most permeable to $\text{K}^{+}$ ions during this phase. They can enter or exit the cell through the ion channels, including the inward-rectifying potassium channel \cite{kusumoto2020ecg}. In pacemaker cells, the membrane potential slowly becomes more positive automatically until it reaches a threshold potential (about $-$40mV) or until it is depolarized by another action potential, coming from a neighboring cell \cite{kusumoto2020ecg,wagner2001marriott}.
    
    \item Phase 0: In this phase, there is a rapid depolarization of the cell membrane. In non-pacemaker cells, the $\text{Na}^{+}$ ion channels open, due to the arrival of an action potential from a neighbouring cell through gap junctions, resulting in a slight increase in the interior cell voltage. If this increased voltage reaches the threshold potential (approximately $-$70 mV) it causes the $\text{Na}^{+}$ channels to open completely \cite{kusumoto2020ecg,wagner2001marriott}. This causes a rapid influx of $\text{Na}^{+}$ ions into the cell, increasing the voltage further to the peak voltage (around $+$50 mV) \cite{kusumoto2020ecg,wagner2001marriott}. If the stimulating action potential is not strong enough to cross the threshold, the $\text{Na}^{+}$ channels will not be open and an action potential will not be produced (All-or-none law) \cite{purves2019neurosciences}. 
    
    \noindent In pacemaker cells, the increase in interior voltage is due to $\text{Ca}^{2+}$ channels which are opened either by it's own depolarization (pacemaker current) or an incoming action potential from a neighbouring pacemaker cell. The rate of depolarization is much slower in pacemaker cells than in non-pacemaker cells. 
    
    \item Phase 1: Only non-pacemaker cells have this phase. There is a rapid closure of the  $\text{Na}^{+}$ ion channels stopping influx of $\text{Na}^{+}$ ions \cite{kusumoto2020ecg}. The $\text{K}^{+}$ ion channels open and close instantaneously resulting in some positive charge inflow into the cells. This phase is characterized by a notch in the waveform of the action potential \cite{santana2010does}.
    
    \item Phase 2: This phase is characterized by a plateau and has the largest duration in the action potential waveform and is also exhibited by only non-pacemaker cells. During this phase, $\text{Ca}^{2+}$ ion channels causing influx of $\text{Ca}^{2+}$ ions and $\text{K}^{+}$ ion channels causing outflux of $\text{K}^{+}$ ions out of the cell results in an almost constant potential difference \cite{kusumoto2020ecg}. 
    
    \item Phase 3: In this phase, the $\text{Ca}^{2+}$ ion channels close and the $\text{K}^{+}$ ion channels open causing rapid repolarization of the cell interior back to its resting potential (around $-$90mV) \cite{kusumoto2020ecg}. Following this, the $\text{K}^{+}$ ion channels close. Both pacemaker and non-pacemaker cells have this phase.
\end{itemize}

\noindent Next, we discuss the refractory period and its two types. The \emph{refractory period}  acts as a safety measure to prevent premature contractions of the heart. The two types of refractory periods are: 
\begin{itemize}
    \item Absolute Refractory Period: This period ranges from the start of phase 0 to the start of phase 3 and lasts for a duration of about 200ms. During this phase, the myocyte is unable to depolarize irrespective of whether the impulse crosses the threshold or not. This is caused due to complete closure or inactivation of the $\text{Na}^{+}$ ion channels since the membrane potential reaches the equilibrium potential of $\text{Na}^{+}$ ions \cite{kusumoto2020ecg}.
    
    \item Relative Refractory Period: This period immediately follows the absolute refractory period and lasts until the end of phase 3 for a duration of about 50ms. During this phase, there is hyperpolarization of the membrane potential which assumes a negative value (refer to Fig.~\pref{fig:AP}). The complete inactivation of the $\text{Na}^{+}$ ion channels stops due to the outflux of some $\text{K}^{+}$ ions. Some $\text{Na}^{+}$ ion channels are also open allowing a possibility of depolarization, provided the impulse is much stronger than the impulse which caused depolarization from phase \cite{kusumoto2020ecg}. 
\end{itemize}

\subsection{Electrocardiogram (ECG/EKG)}

The net electrical activity of the heart is recorded by the ECG. Since the atrial and ventricular action potential amplitudes are much larger than those of the individual pacemaker cells, the ECG can be regarded as the electrical activity of the atria and the ventricles. Conventionally, the ECG is a 12-lead device (10 electrodes - 9 chest and limb leads and one exploratory), where the total magnitude of the electrical potential of heart is measured from twelve different angles (leads are equivalent to electrodes) placed on the limbs and the chest. A positive depolarization wave travelling towards the positive lead (exploratory lead is the positive electrode whereas the others are reference electrodes) is recorded as positive whereas a negative depolarization wave travelling away from the positive lead is recorded as negative \cite{kusumoto2020ecg}. In this paper, we are mainly concerned with lead II which measures the electrode potential difference between the left leg (exploratory) and the right arm. The heart is observed from an angle of 60$\degree$ from the frontal plane. \\

\begin{figure}[H]
    \begin{subfigure}{.5\textwidth}
        \centering
        \includegraphics[scale=0.3]{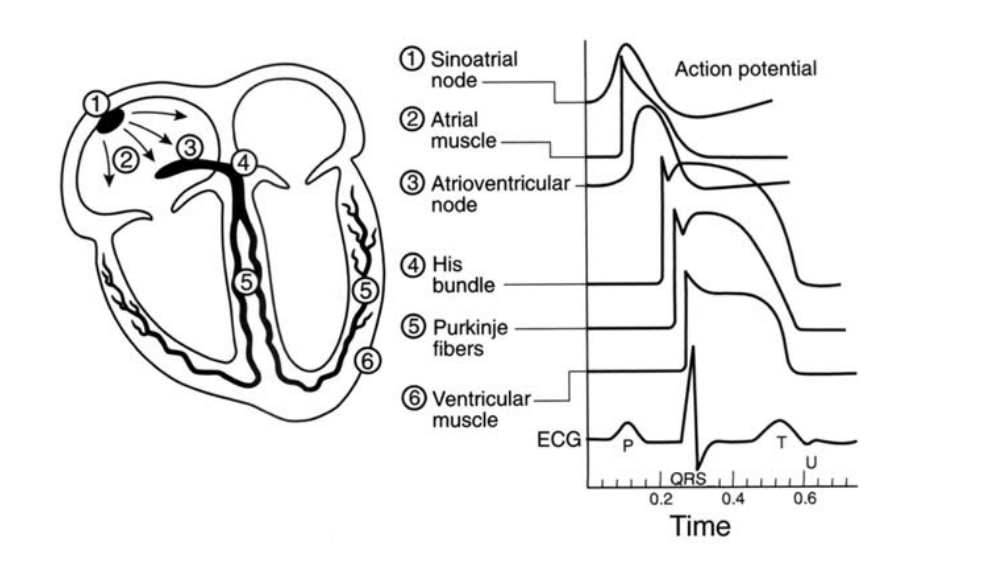}
        \caption{}
        \label{fig:ecg_gen}
    \end{subfigure}
    \begin{subfigure}{.5\textwidth}
        \centering
        \includegraphics[scale=0.2]{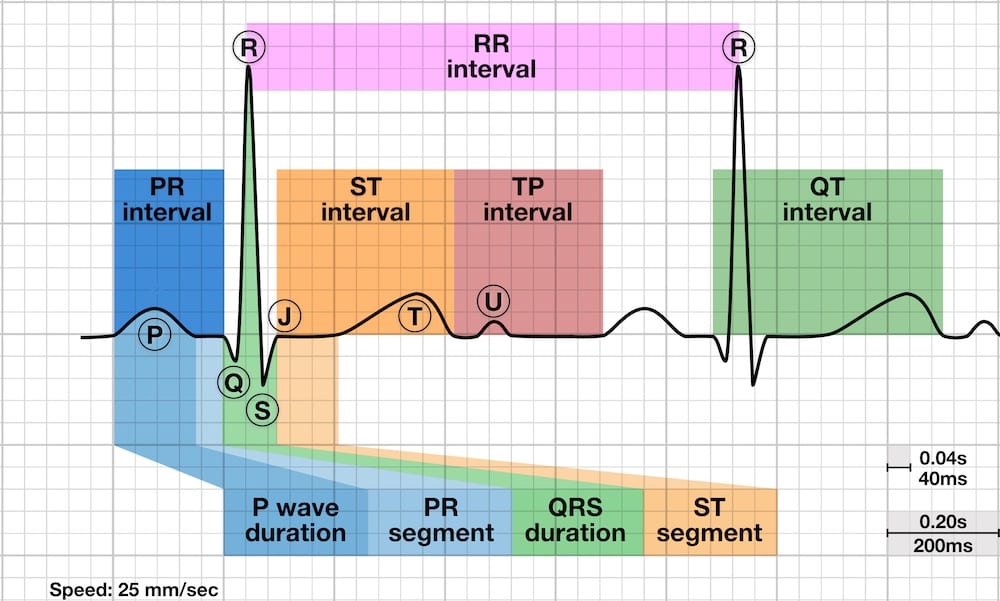}
        \caption{}
        \label{fig:ecg_components}
    \end{subfigure}
    \caption{ECG terminology. (a) Generation of the normal sinus rhythm ECG \cite{kusumoto2020ecg}, and (b) Normal Sinus Rhythm ECG intervals, waves and segments \cite{litfl}}
 \end{figure}

\noindent The main wave components of the ECG waveform are - the P wave, QRS complex and the T wave (refer to Fig.~(\pref{fig:ecg_gen})). Since the ventricles have larger size, the ventricular contribution to the ECG waveform will be of larger amplitude than the atrial waveforms. The default heart rhythm is called the normal sinus rhythm where the action potentials generate from the SA node. These are then forwarded to the AV node causing synchronized contraction of atrial muscles and then forwarded to the His-Purkinje complex causing the ventricular muscles to contract. The features of normal sinus rhythm are (i) a heart rate of 60-100 bpm  in human adults where each depolarization and repolarization cycle (ii) upright P wave, followed by the QRS complex and T wave (in lead II) (iii) a constant PR interval and (iv) the duration of the QRS complex, $\lesssim$ 100ms.
 
The ECG intervals, waves and segments of a normal sinus rhythm ECG are discussed below (refer to Fig.~\pref{fig:ecg_gen} and Fig.~\pref{fig:ecg_components}):
\begin{itemize}
    \item P-Wave: It represents atrial depolarization and is the first positive deflection in the ECG (lead II). It has a normal duration of less than 0.12 s. The first one-third of the P wave represents to right atrial activation, the last one-third corresponds to left atrial activation while the middle one-third is a combination of the two atrial activation \cite{hampton2019ecg,wagner2001marriott}.
    
    \item PR-Interval: This represents the time from the onset of the P wave to the start of the QRS complex. It corresponds to the conduction of the action potential through the AV node. The normal PR interval is between 120$–$200 ms.
    
    \item PR-Segment: In normal cases, it is a flat, isoelectric segment between the end of the P wave and the start of the QRS complex. PR segment depression is observed in acute pericarditis and atrial ischaemia can include both elevation and depression \cite{litfl}.
    
    \item QRS-Complex: The QRS complex is composed of three different waves. The first one is the Q-Wave, which is a small negative deflection that precedes the R-Wave and represents the left-to-right depolarization of the interventricular septum. The next one is the R-wave, which is the first large upward deflection after the P wave, representing early ventricular depolarization. The third and the last one is the S-wave, which is a small negative deflection representing the final depolarization of the ventricles at the base of the heart. The normal QRS-interval width is usually 70$-$100 ms. However, a duration of 110 ms is sometimes also observed in healthy subjects \cite{litfl,hampton2019ecg,wagner2001marriott}. 
    
    \item QT-Interval: It is the interval in time from the start of the QRS-complex to the end of the T-wave, representing the time duration for ventricular depolarization and repolarization. This is short for faster heart rates (sinus tachycardia) and long for slower heart rates (sinus bradycardia) \cite{litfl}.

    \item RR-Interval: It is the time between two successive R-Waves of the QRS-complex. This interval is useful in calculating the heart rate and the QT-interval using specialized formulae which are applicable for varying heart rates. 
    
    \item ST-Segment: It is a flat, isoelectric section of the ECG between the end of the S wave and the beginning of the T wave, representing the interval between ventricular depolarization and repolarization. ST segment elevation or depression is mainly observed in myocardial ischemia or infraction \cite{litfl}. 
\end{itemize}

\noindent The discussion in this current section provides a description of the cardiac conduction system whose electrical activity will be studied using an ODE-based model in the following section. These timescales and intervals of normal sinus rhythm are of prime importance while fitting an ECG data with optimization algorithms. One of the prime aims of this model is to be able to separately represent the spontaneous depolarization of the pacemaker cells and the stimulus driven depolarization and repolarization of the non-pacemaker cells of the atria and ventricles.


\section{Formulation of ODE based model}
\label{Formulation}

In this section, we discuss the ODE-based model for the cardiac pacemaker and non-pacemaker cardiomyocytes. The depolarization and repolarization waves in the cardiac myocytes, produced by the pacemaker and the non-pacemaker cells of the atria and ventricles, are modeled using the modified van der Pol (VDP) equations \cite{van1928lxxii,grudzinski2004modeling} and the FitzHugh-Nagumo (FHN) equations for excitable media \cite{fitzhugh1961impulses} respectively. Ryzhii \textit{et al.} \cite{ryzhii2014heterogeneous} connected these two models to develop an ODE-based cardiac conduction model. In this work, we have made further modifications to this model.

\subsection{Modified van der Pol Equation (VDP)}

 The action potentials that are generated by the cardiac pacemaker cells of the Sinoatrial (SA), Atrioventricular (AV) nodes and the His-Purkinje (HP) complex cause contraction in the excitable non-pacemaker cells of the atria and ventricles. To model such self-sustaining pacemakers, the van der Pol oscillator is used which is modified in such a way that its phase space resembles that of a neuron. It also has parameters that allow for the change in the refractory period, frequency of oscillations, amplitude and the resting potential \cite{grudzinski2004modeling,morris1981voltage}.\\
 
 \noindent The modification to the ordinary van der Pol oscillator is introduced by replacing the harmonic force term with a cubic Duffing oscillator term \cite{morris1981voltage} and by including an anharmonic damping term in the ordinary van der Pol equation  \cite{grudzinski2004modeling}. The modified equation is given by,
\begin{equation}
    \ddot{x}+\alpha (x-v_{1})(x-v_{2})\dot{x}+\frac{x(x+d)(x+e)}{(d\,e)}=0
\end{equation}
where $x$ is the transmembrane potential of the cell and the parameters satisfying $\alpha,\mu,d,e > 0$, $v_{1}v_{2}<0$. \\

\noindent This system can be further decomposed into two first order systems where the variables $x$ and $y$ are the transmembrane potential is its derivative respectively.
\begin{align}
   \label{mVDP_1storder}
    \dot{x}&=y  \notag \\
    \dot{y}&=-\alpha (x-v_{1})(x-v_{2})y-\frac{x(x+d)(x+e)}{(d\,e)}
\end{align}
The fixed points of Eq.~\pref{mVDP_1storder} are: $(0,0)$, $(-d,0)$ and $(-e,0)$ which are an unstable focus, a saddle point and a stable node respectively.
The main features of the phase space and parameter dependencies have been discussed in detail by Grudzi\'nski \textit{et al.} \cite{grudzinski2004modeling}. 


\subsection{FitzHugh-Nagumo Equation (FHN)}
The cardiac myocytes do not possess self-oscillatory behaviour. They must be stimulated by some external electrical impulse to produce a signal. In this case, the external electrical impulse is provided by the action potential of the pacemaker cells. These types of excitable media can be modeled using the VDP equations, as well as the FHN equations which is a 2D analogue of the more general Hodgkin-Huxley (HH) equations. The model involves:
\begin{itemize}
    \item The transmembrane potential, $z$, which has a resting value and starts increasing when the external signal is high enough to cross a threshold value. Next it rapidly increases (depolarization) until it reaches a peak value and  then suddenly decreases (repolarization) to its resting value.
    \item The blocking mechanism, $v$, which brings the transmembrane potential back to its resting potential.
\end{itemize} 
The equations of interest are:
\begin{align}
    \label{Eqn:General_FHN_1}
    \dot{z}&=-cz(z-w_{1})(z-w_{2})-bv-dvz+I \notag \\
    \dot{v}&=h(z-gv)
\end{align}
\noindent Notice that unlike the modified van der Pol equations, here the product of $w_{1}$ and $w_{2}$ must be positive ($w_{1}w_{2}>0$) and $w_{1}<w_{2}$, so as to eliminate the self-oscillatory nature. \\

\subsection{Ryzhii-FHN Model}
\begin{figure}[H]
   \centering
    \includegraphics[scale=0.06]{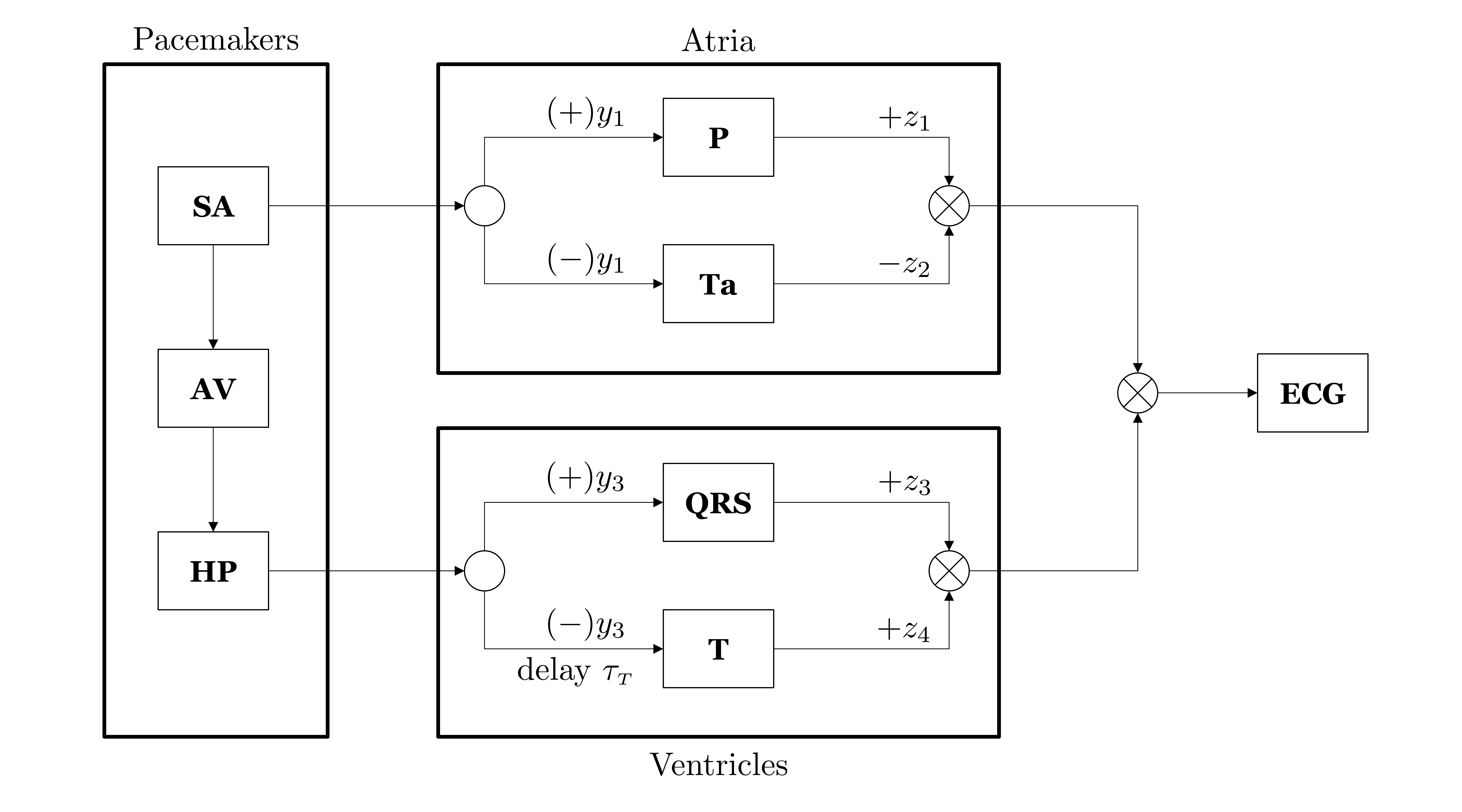}
    \caption{Block Diagram of the Cardiac Conduction System }
    \label{blockdiagram}
\end{figure}

We present a schematic block diagram of the cardiac conduction system (refer to Fig.~\pref{blockdiagram}), inspired by Ryzhii \textit{et al.,} \cite{ryzhii2014heterogeneous}. It is noted here that several authors have modeled this system with both position and velocity coupling allowing for bidirectional signal flow. They have observed that the cardiac pacemaking system can be reproduced, up to 200 bpm, with pure velocity coupling and unidirectional flow \cite{di1998model,west1985nonlinear}. Thus, to achieve synchronized interaction between the three pacemaker blocks, we have used unidirectional velocity coupling, as used in Ryzhii \textit{et al.,} \cite{ryzhii2014heterogeneous}.\\ 

\noindent The heart consists of three pacemaker complexes: SA, AV and HP. These pacemaker complexes are separately modeled by using Eq.~\pref{mVDP_1storder}. As discussed earlier, AV node has a unidirectional velocity coupling with SA node and the HP node has a unidirectional coupling with the AV node (refer to Fig.~\pref{blockdiagram}). The VDP equations corresponding to SA, AV and HP are described by Eq.~\pref{eq:SA_eqn}, Eq.~\pref{eq:AV_eqn} and Eq.~\pref{eq:HP_eqn} respectively:
\begin{align}
    \label{eq:SA_eqn}
    \dot{x_{1}}&=y_{1} \notag\\ 
    \dot{y_{1}}&=-a_{1}y_{1}(x_{1}-u_{11})(x_{1}-u_{12})-f_{1}x_{1}(x_{1}+d_{1})(x_{1}+e_{1})
\end{align}
\begin{align}
    \label{eq:AV_eqn}
    \dot{x_{2}}&=y_{2} \notag \\
    \dot{y_{2}}&=-a_{2}y_{2}(x_{2}-u_{21})(x_{2}-u_{22})-f_{2}x_{2}(x_{2}+d_{2})(x_{2}+e_{2})\notag\\&+K_{\scriptscriptstyle SA-AV}(y_{1}^{\tau_{\scriptscriptstyle SA-AV}}-y_{2})
\end{align}
\begin{align}
    \label{eq:HP_eqn}
    \dot{x_{3}}&=y_{3} \notag\\
    \dot{y_{3}}&=-a_{3}y_{3}(x_{3}-u_{31})(x_{3}-u_{32})-f_{3}x_{3}(x_{3}+d_{3})(x_{3}+e_{3})\notag\\&+K_{ \scriptscriptstyle AV-HP}(y_{2}^{\tau_{\scriptscriptstyle AV-HP}}-y_{3})
\end{align}

Here $y_{i}^{\tau_{n}}=y_{i}(t-\tau_{n})$ and $\tau_{n}$ are the corresponding time delays. 

\begin{figure}[H]
    \begin{subfigure}{.5\textwidth}
        \centering
        \includegraphics[scale=0.6]{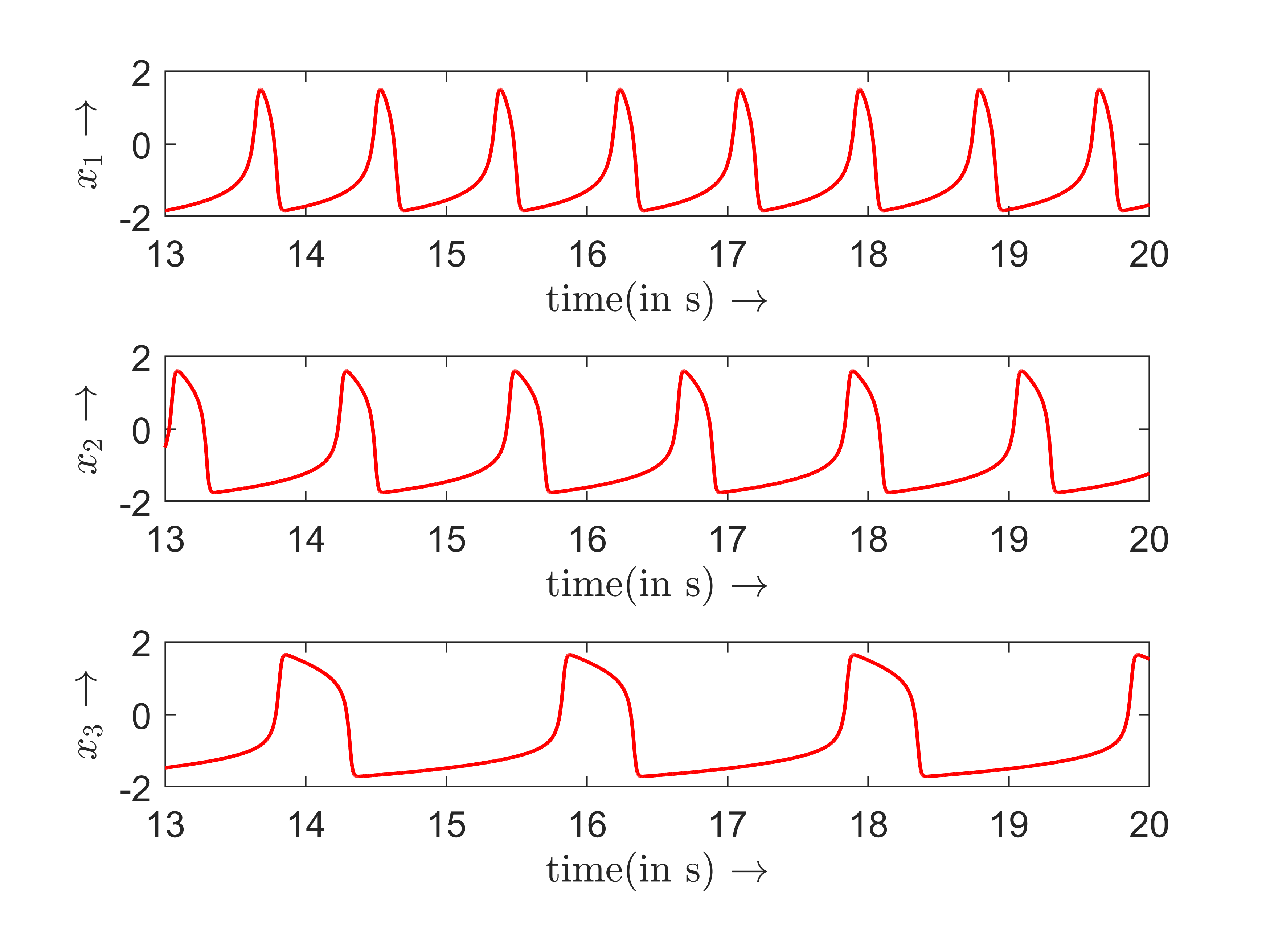}
        \caption{}
    \end{subfigure}
    \begin{subfigure}{.5\textwidth}
        \centering
        \includegraphics[scale=0.6]{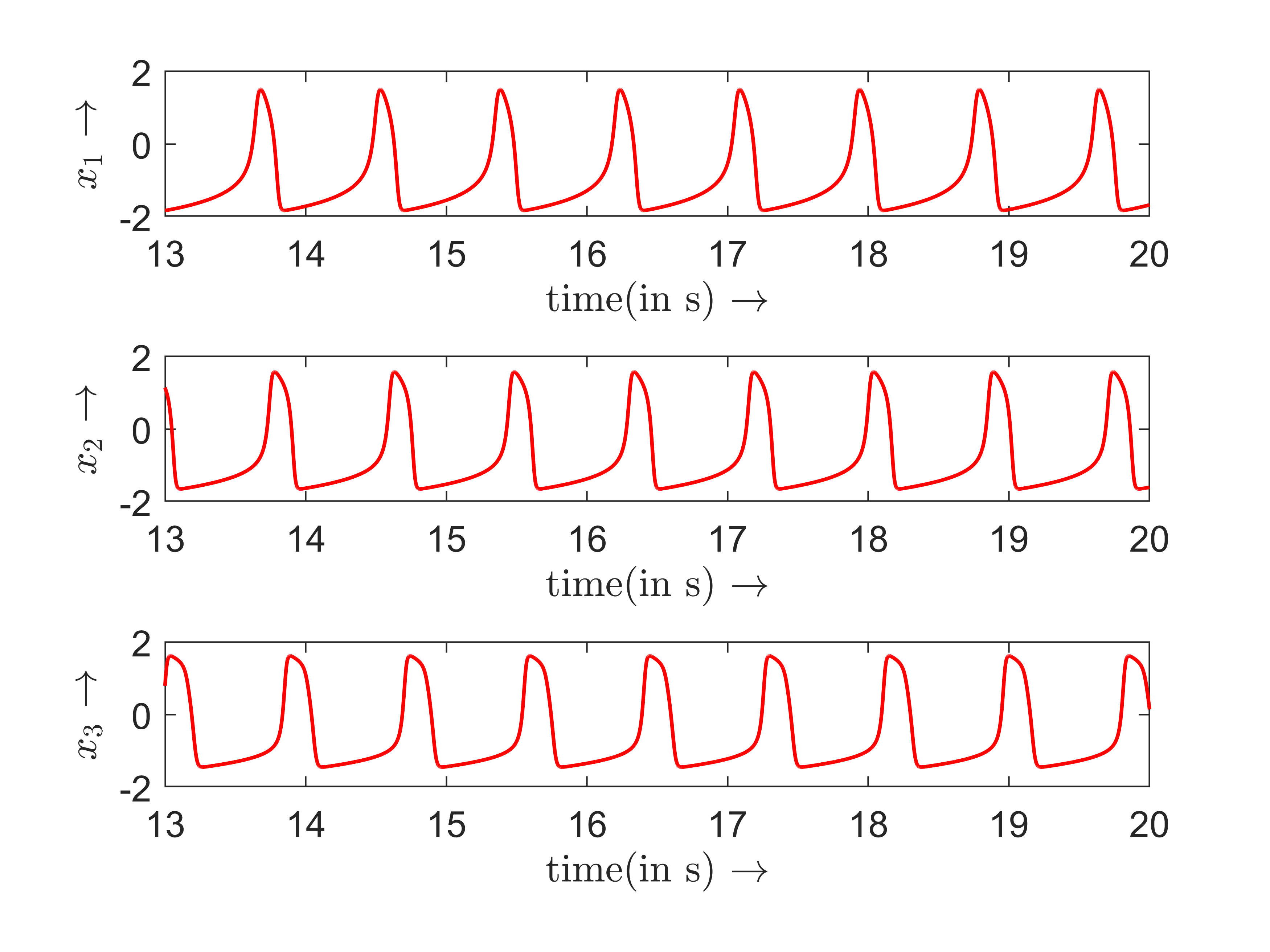}
        \caption{}
    \end{subfigure}
    \caption{Action potentials of pacemaker systems. (a) Uncoupled pacemaker system, and (b) Coupled pacemaker system}
    \label{fig:vdp_osc}
 \end{figure}

\noindent The above action potentials generated by the pacemaker complex act as stimuli for the non-pacemaker cells of the atria and ventricles. An ECG is generated by the superposition of the atrial signal and the ventricular signal (Fig.~\pref{blockdiagram}). The atrial signal is composed of the atrial depolarization and repolarization waves which are called the P and Ta waves respectively. The ventricular signal is composed of ventricular depolarization and repolarization waves which are called the QRS and T waves respectively. These four waves vary in shape as viewed from Fig.~\pref{fig:ecg_components}. We need a case-by-case modification of the above equation (Eq.~\pref{Eqn:General_FHN_1}) to construct each of the waves and finally superpose them together to get the ECG signal, resembling the standard Einthoven lead II signal. The pacemaker action potentials have negligible amplitude and thus their presence on the ECG signal can be ignored. The major contribution that these potentials have is in driving the oscillation in the non-pacemaker cells of the atria and the ventricles. These waves, along with the self-sustained response of the pacemaker cells, are used to generate the synthetic ECG signal. The sets of equations for the waves are :

\begin{align}
\label{Eqn:P_Wave}
    \dot{z_{1}}=&-k_{1}(c_{1}(z_{1}-w_{11})(z_{1}-w_{12})-b_{1}v_{1}-d_{1}v_{1}z_{1}+I_{\scriptscriptstyle AT_{\scriptscriptstyle De}}) \nonumber \\ 
    \dot{v_{1}}=&k_{1}h_{1}(z_{1}-g_{1}v_{1}) \\
    & \nonumber \\ 
\text{where, } I_{\scriptscriptstyle AT_{\scriptscriptstyle De}}=&\begin{cases}
0 & ;y_{1}\leq 0 \nonumber \\
K_{\scriptscriptstyle AT_{\scriptscriptstyle De}}y_{1} & ;y_{1}>0
\end{cases}
\end{align}

\begin{align}
\label{Eqn:Ta_Wave}
   \dot{z_{2}}=&-k_{2}(c_{2}(z_{2}-w_{21})(z_{2}-w_{22})-b_{2}v_{2}-d_{2}v_{2}z_{2}+I_{\scriptscriptstyle AT_{\scriptscriptstyle Re}}) \nonumber \\ 
   \dot{v_{2}}=&k_{2}h_{2}(z_{2}-g_{2}v_{2}) \\
    & \nonumber \\
\text{where, } I_{\scriptscriptstyle AT_{\scriptscriptstyle Re}}=&\begin{cases}
 -K_{\scriptscriptstyle AT_{\scriptscriptstyle Re}}y_{1} & ;y_{1}\leq 0 \nonumber \\
 0 & ;y_{1}>0
\end{cases}
\end{align}

\begin{align}
    \label{Eqn:QRS_Wave}
    \dot{z_{3}}=&-k_{3}(c_{3}(z_{3}-w_{31})(z_{3}-w_{32})-b_{3}v_{3}-d_{3}v_{3}z_{3}+I_{\scriptscriptstyle VN_{\scriptscriptstyle De}}) \nonumber \\ 
    \dot{v_{3}}=&k_{3}h_{3}(z_{3}-g_{3}v_{3}) \\
    & \nonumber \\
 \text{where, } I_{\scriptscriptstyle VN_{\scriptscriptstyle De}}=&\begin{cases}
0 & ;y_{3}\leq 0 \nonumber \\
K_{\scriptscriptstyle VN_{\scriptscriptstyle De}}y_{3} & ;y_{3}>0
\end{cases}
\end{align}

\begin{align}
\label{Eqn:T_Wave}
    \dot{z_{4}}=&-k_{4}(c_{4}(z_{4}-w_{41})(z_{4}-w_{42})-b_{4}v_{4}-d_{4}v_{4}z_{4}+I_{VN_{Re}}) \nonumber \\ 
    \dot{v_{4}}=&k_{4}h_{4}(z_{4}-g_{4}v_{4}) \\
    & \nonumber \\
    \text{where, } I_{\scriptscriptstyle VN_{\scriptscriptstyle Re}}=&\begin{cases}
 -K_{\scriptscriptstyle VN_{\scriptscriptstyle Re}}y_{3}(t-\tau_{\scriptscriptstyle T}) & ;y_{3}\leq 0 \nonumber \\
 0 & ;y_{3}>0
\end{cases}
\end{align}

\noindent Eq.~\pref{Eqn:P_Wave},~\pref{Eqn:Ta_Wave},~\pref{Eqn:QRS_Wave} and~\pref{Eqn:T_Wave} represents the P Wave (atrial depolarization), Ta Wave (atrial repolarisation), QRS Wave (ventricular depolarization) and the T Wave (ventricular repolarisation) respectively. We have introduced a new parameter $\tau_{\scriptscriptstyle T}$ whose reason of inclusion will be discussed shortly.

\subsection{$\tau_{\scriptscriptstyle T}$-Ryzhii Model}
In this subsection, we introduce some critical changes in the earlier established FHN model by Ryzhii \textit{et al.,} \cite{ryzhii2014heterogeneous} to reconcile with the mismatch that was observed while fitting the Ryzhii FHN model with real ECG data and the earlier model. 
The problem of the model without the time lag $\tau_{\scriptscriptstyle T}$ in the FHN equation of the T-Wave is that the peak-to-peak Q and T-Wave interval has a mismatch with the actual ECG data. The peak positions were fixed once all the parameters were fitted. With the inclusion of lag $\tau_{\scriptscriptstyle T}$, the T-Wave peak could be shifted efficiently to its desired position.

\begin{figure}[H]
    \centering
    \includegraphics[scale=0.6]{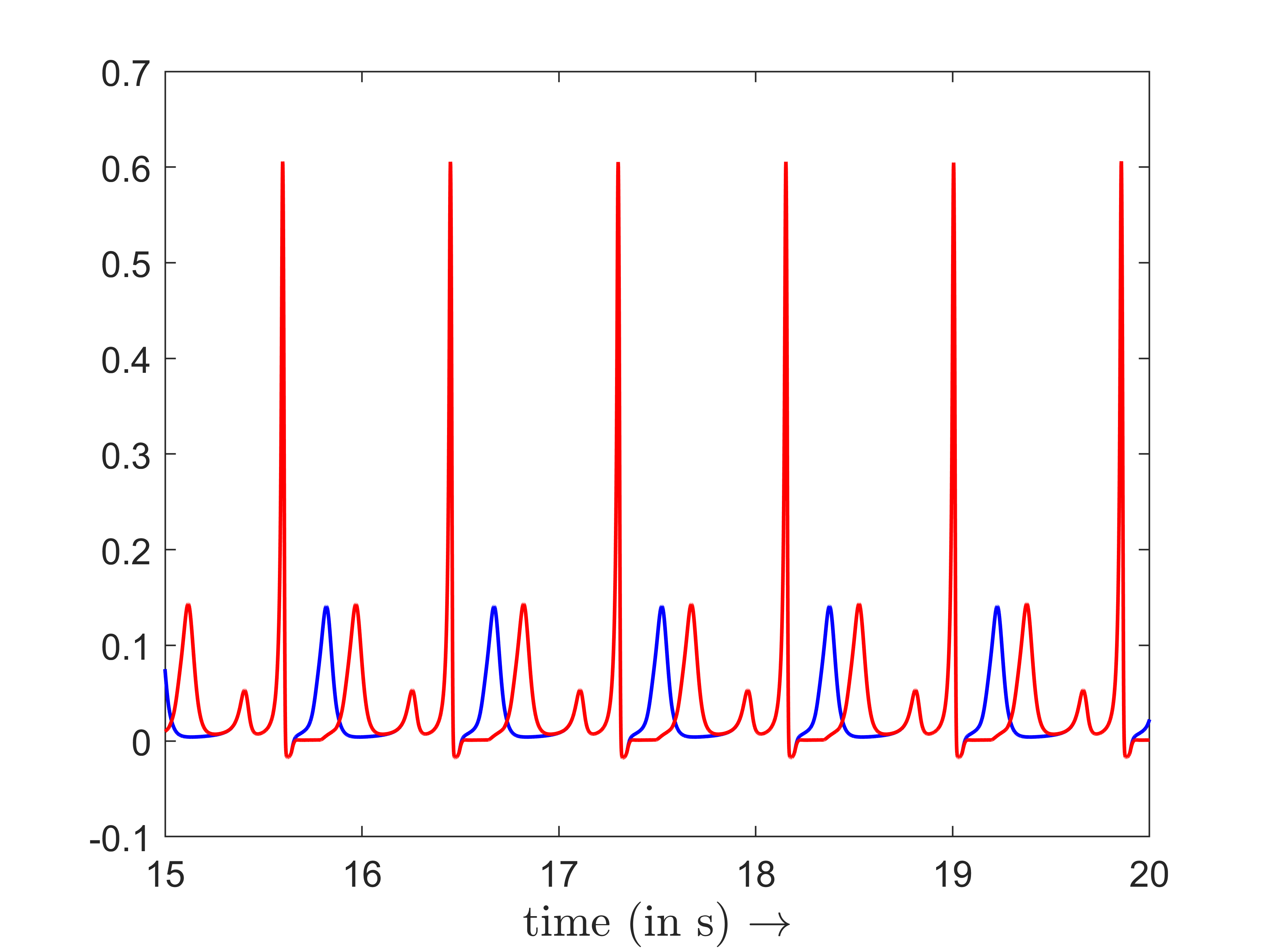}
    \caption{Change in ST interval (in s) without (solid blue line) and with lag $\tau_{\scriptscriptstyle T}=0.15$ s (solid red line)}
    \label{fig:tau_T_lag}
\end{figure}

The parameters used in the above Fig.~\pref{fig:tau_T_lag} are the ones used by Ryzhii \textit{et al.,} \cite{ryzhii2014heterogeneous} which are:
\begin{itemize}
    \item Pacemaker complex parameters: $a_{1}=40, u_{11}=-u_{12}=0.83, f_{1}=22, d_{1}=3, e_{1}=3.5, a_{2}, u_{21}=-u_{22}=0.83, f_{2}=8.4, d_{2}=3, e_{2}=5, a_{3}=50, u_{31}=-u_{32}=0.83, f_{3}=1.5, d_{3}=3, e_{3}=12.$
    \item Atrial and Ventricular wave parameters: $k_{1}=2000, c_{1}=0.26, w_{11}=0.13, w_{12}=1, h_{1}=0.004, g_{1}=1, K_{\scriptscriptstyle AT_{\scriptscriptstyle De}}=4 \times 10^{-5}, d_{11}=0.4, k_{3}=10^{4}, c_{3}=0.12, w_{31}=0.12, w_{32}=1.1, b_{3}=0.015, h_{3}=0.008, g_{3}=1, K_{\scriptscriptstyle VN_{\scriptscriptstyle De}}=9 \times 10^{-5}, d_{31}=0.09, k_{4}=2000, c_{4}=0.1, w_{41}=0.22, w_{42}=0.8, h_{4}=0.008, g_{4}=1, K_{\scriptscriptstyle VN_{\scriptscriptstyle Re}}=6 \times 10^{-5}, d_{41}=0.1 .$
\end{itemize}


\subsection{Effect of delays on the ECG}

\begin{figure}[H]
        \centering
        \includegraphics[scale=0.35]{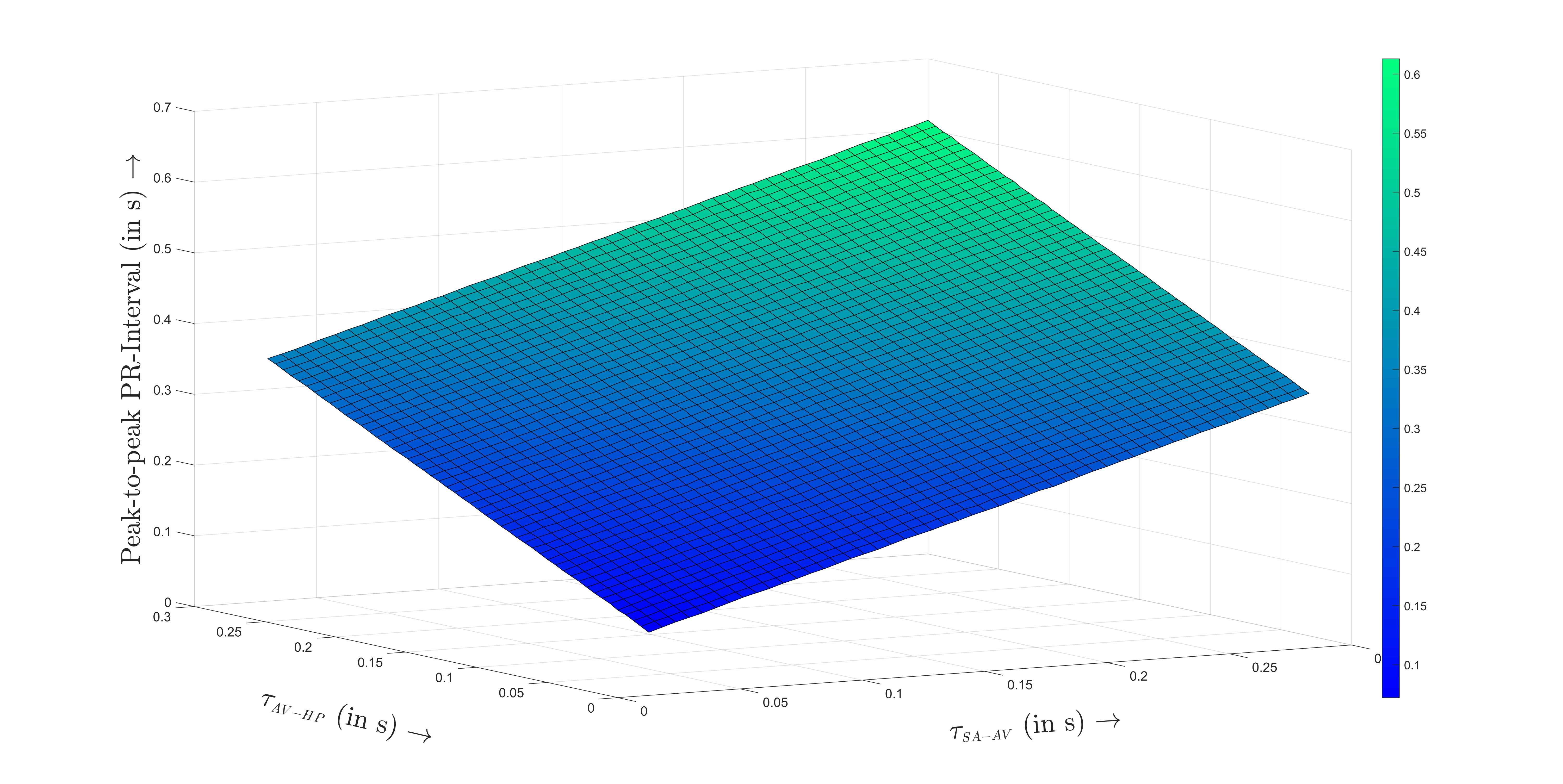}
        \caption{Variation of peak-to-peak PR-Interval with $\tau_{\scriptscriptstyle SA-AV}$ and $\tau_{\scriptscriptstyle AV-HP}$}
        \label{delay_PR}
 \end{figure}

This model has three delays, $\tau_{\scriptscriptstyle SA-AV}$, $\tau_{\scriptscriptstyle AV-HP}$ and $\tau_{\scriptscriptstyle T}$. The effects of these delays are discussed here. The delays, $\tau_{\scriptscriptstyle SA-AV}$ and $\tau_{\scriptscriptstyle AV-HP}$ elongate the duration of the PR interval of the ECG. Also, the effect of these delays on the PR interval is additive. This means the PR interval depends on the $\tau_{\scriptscriptstyle SA-AV}+\tau_{\scriptscriptstyle AV-HP}$, which is demonstrated in Fig.~\pref{delay_PR}.

\begin{figure}[H]   
    \includegraphics[scale=0.38]{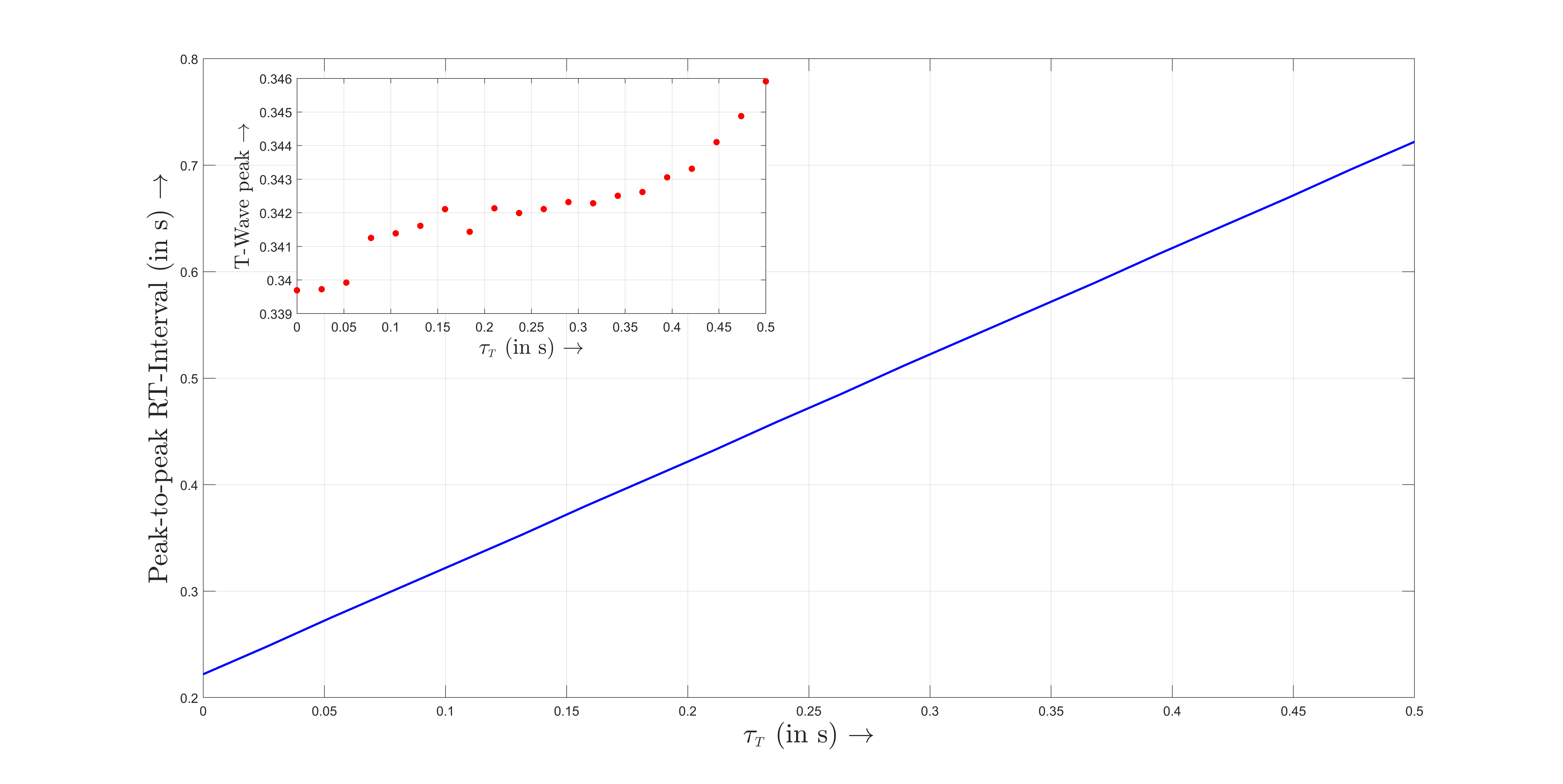}
    \caption{Variation of peak-to-peak RT-Interval with $\tau_{\scriptscriptstyle T}$ and variation of T-Wave peak with $\tau_{\scriptscriptstyle T}$ (inset)}
    \label{fig:model_lags}
\end{figure}

\noindent The delay, $\tau_{\scriptscriptstyle T}$ increases the RT interval as mentioned in the previous subsection. This feature is represented in Fig.~\pref{fig:model_lags}. The $\tau_{\scriptscriptstyle T}$ also increases the peak of the T-wave which is shown at the inset of Fig.~\pref{fig:model_lags}. However, this increment is very small and can be neglected. 

\section{Model Fitting} \label{Mod_fit}
This section discusses the step-by-step process to fit the ECG data by the above-described ODE-based model. The data for the ECG of different physical conditions are taken from several data repositories and scientific articles. 
\subsection{Feature extraction}
In this subsection, we have discussed the features that have been extracted from the ECG data and the method of extracting them. An ECG data can have fluctuating frequency and isoelectric baseline. Also, the magnitude of the peaks of the P, R, and T waves can vary with time. Thus it is difficult to fit this model directly to the ECG data. Hence we have opted to optimize our model by using the different features of the ECG data. Also, it is easier for detection algorithms to pick up the peaks of the data than any other general data point. Thus to maintain simplicity in our algorithm, we have used the peak-to-peak intervals as our basis, for example, the PR interval refers here as the interval between the peaks of the P-wave and the QRS-complex. The features that are estimated from the data are listed below:
\begin{itemize}
    \item Time period $T$: Average time period between two successive waves in the ECG signal.
    \item PR interval: Average peak-to-peak interval between the P-wave and the R-wave.
    \item RT interval: Average peak-to-peak interval between the R-wave and the T-wave.
    \item Average peak height of the P, QRS, and T waves. 
    \item Average width of the P waves at 50\% and 10\% height of the peaks.
    \item Average width of the QRS waves at 50\% and 10\% height of the peaks.
    \item Average width of the T waves at 50\% and 10\% height of the peaks.
    \item Average time interval between the position of the peak of the T-wave and the position to the left where the height of the T-wave is 50\%  of the peak height.
\end{itemize}

\noindent All these estimates represent the average characteristic of ECG data. To maintain proper synchronization between different waves (P, T, and QRS), it is essential to estimate the time period, PR interval, and RT interval of the ECG signal. Measurements of peak heights and widths for P, T, and QRS waves help us to maintain the shape of the ECG signal.   

\subsection{Parameter Selection}

This ODE-based model consists of a total of seven sets of equations where three sets are modified van der Pol (VDP) equations, representing the action potential of the cardiac pacemaker cells and the other four sets of equations are modified FHN equations which are used for modeling the atrial and ventricular depolarization and repolarization. These seven sets of equations have over sixty different parameters. Thus the selection of important parameters is essential.
\\~\\
The action potential of SA, AV and HP nodes are modeled with three sets of coupled modified van der Pol equations. The van der Pol equation corresponding to the SA node (Eq.~\pref{eq:SA_eqn}) is very important as it controls the heart rate. For normal sinus rhythm, the frequencies of AV and HP nodes are also controlled by the SA node due to strong coupling. Thus all parameters of the VDP oscillator for the SA node are very important. However, we have assumed that the absolute of the two threshold values of this equation, $u_{11}$ and $u_{12}$ are equal, $u_{11}=-u_{12}=u_{1}$ (say).
\\~\\
The van der Pol equations for the AV and HP nodes (Eq.~\pref{eq:AV_eqn} and \pref{eq:HP_eqn}) are strongly coupled with the VDP oscillator for the SA node (for normal sinus rhythm). Therefore the original features of the VDP oscillator for the AV and HP nodes are overshadowed by the VDP oscillator for the SA node and the behaviour of these coupled equations does not change much by changing the parameters (except coupling constants and delays) of it. The coupling constants and the time delays are of great importance as they control the synchronization of the ECG waves. Thus we have excluded the parameters of these equations other than the coupling constants and the time delays from the list of parameters whose best fit values have been determined. The values of the parameters which have been fixed \textit{a priori} are chosen to be the values used by Ryzhii \textit{et. al.} \cite{ryzhii2014heterogeneous}.
\\~\\
Each set of FHN equation models different components of ECG signal like P wave, Ta wave, QRS complex, and T wave. The Ta wave is absorbed by the QRS complex and cannot be detected separately. Thus we have opted to exclude this FHN equation from fitting. The other three sets of FHN equations and their parameters are essential for fitting. However, for P-wave and T-wave the hyper-polarization of FHN should be neglected, and thus the parameters, $b_{1}$ and $b_{4}$ are set to zero. The table given below (Tab.~\pref{sel_param_tab}) shows the full list of the parameters which are fitted for this ODE-based cardiac model. The process of optimizing these parameters is discussed in the next subsection. 

\begin{table}[H]
    \centering
    \begin{tabular}{|c|c|c|}
        \hline
        Component & Model used & Parameters for fitting \\
        \hline
        SA node & Modified van der Pol equation & $a_{1}$, $u_{11}=-u_{12}=u_{1}$, $f_{1}$, $d_{1}$, and $e_{1}$ \\
        AV node & Modified van der Pol equation with SA coupling & $K_{SA-AV}$ and $\tau_{SA-AV}$ \\
        HP node & Modified van der Pol equation with AV coupling & $K_{AV-HP}$ and $\tau_{AV-HP}$ \\
        P-wave & Modified FHN equation & $k_{1}$, $c_{1}$, $w_{11}$, $w_{12}$, $d_{1}$, $h_{1}$, $g_{1}$, and $K_{AT_{De}}$.\\
        Ta-wave & Modified FHN equation & --- \\
        QRS-complex & Modified FHN equation & $k_{3}$, $c_{3}$, $w_{31}$, $w_{32}$, $b_{3}$, $d_{3}$, $h_{3}$, $g_{3}$, and $K_{VN_{De}}$.\\
        T-wave & Modified FHN equation & $k_{4}$, $c_{4}$, $w_{41}$, $w_{42}$, $d_{4}$, $h_{4}$, $g_{4}$, and $K_{VN_{Re}}$.\\ 
        \hline
    \end{tabular}
    \caption{List of selected parameters for fitting.}
    \label{sel_param_tab}
\end{table}

\subsection{Optimization process}
\begin{figure}[H]
    \centering
    \includegraphics[scale=0.38]{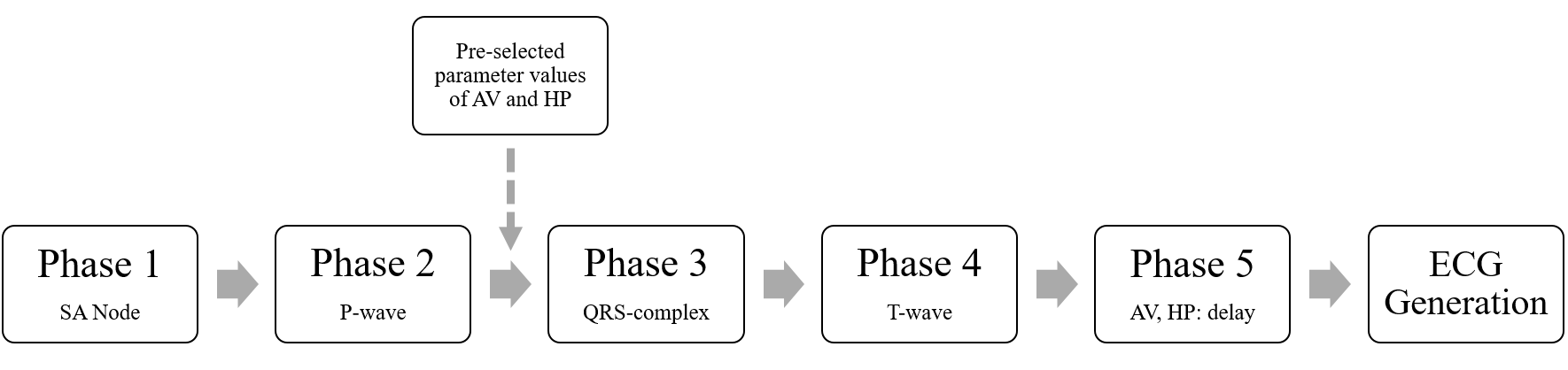}
    \caption{Schematic diagram of the optimization process.}
    \label{schem_GA}
\end{figure}

In this subsection, we have discussed the optimization process for the ODE based electrophysiological model. Although we have judiciously selected the important parameters, there is still a large number of parameters for fitting. Therefore we have divided the optimization process into multiple phases. The schematic diagram of the optimization process is shown in Fig.~\pref{schem_GA}.

The description of each phase is given below:

\begin{itemize}
    \item Phase 1: In this phase the parameters of the VDP oscillator for the SA node are optimized. As mentioned earlier, the SA node is the main generator of the ECG signal. Additionally, the VDP equation corresponding to this is independent of other equations and thus can be optimized separately. The ECG signal does not give much information about the action potential generated by the pacemaker cells of the SA node. However, the frequency or the time period ($T$) of the ECG signal can be estimated. According to this model, the P wave and Ta wave are generated by the positive and negative peaks of the velocity component of the van der Pol equation of the SA node respectively. Thus it can be assumed that the peak-to-peak interval between the P and Ta waves is approximately equal to the time interval between the positive peak and the negative peak ($\text{SA}_{+-}$) of the velocity component of the VDP oscillator for the SA node i.e., P-Ta interval $\approx$ $\text{SA}_{+-}$. However, Ta waves cannot be found in an ECG signal generally and thus it is further approximated that P-Ta interval $\approx$ P-R interval and consequently $\text{PR interval} \approx \text{SA}_{+-}$. 

    Hence, the parameters of the VDP oscillator for the SA node are optimized with the array
    $\left\{T,T_{PR}\right\}$. The optimization problem for Phase 1 can be written as,

    \begin{equation}
        \text{Minimize,}~f(X)=\left\{w_{1}\left(T-\hat{T} \right)\right\}^{2}+\Bigl\{w_{2}\Bigl(T_{PR}-\text{SA}_{+-}\Bigr)\Bigr\}^{2}
    \end{equation}

    \begin{equation*}
        \text{where,}~X=\left\{a_{1}, u_{1}, f_{1}, d_{1}, e_{1}\right\}.
    \end{equation*}

    Here, $f(X)$= Objective function, $X$= Set of decision variables, and $\hat{T}$= Time period of  VDP oscillator for the SA node. Also, $w_{1}$ and $w_{2}$ are the weights that have been chosen to scale the two components of the objective function equally.
    \item Phase 2: In Phase 2, the FHN equation for the P wave is optimized. The stimulus $I(t)$ of this FHN equation depends on the positive peaks of the velocity component of the VDP oscillator for the SA node Eq.~\pref{Eqn:P_Wave}. This stimulus $I(t)$ has been estimated (without scaling factor $K_{AT_{De}}$) using the optimized parameter values from Phase 1. 
    
    The main features of the P waves that are extracted from an ECG signal are (1) Average Peak height of the P waves ($P_{P}$), (2) Average width of the P waves at 50\% height of the peaks ($T_{P_{50}}$), and (3) Average width of the P waves at 10\% height of the peaks ($T_{P_{10}}$). Also, the average time period ($T$) is also necessary for proper synchronization. Thus the array of values that are used for optimizing the parameters of the FHN equation of the P wave is $\left\{T, P_{P}, T_{P_{50}}, T_{P_{10}}\right\}$. The optimization problem for Phase 2 can be written as:

    \begin{equation}
        \text{Minimize,}~f(X)=\left\{w_{1}\left(T-\hat{T} \right)\right\}^{2}+\left\{w_{2}\left(P_{P}-\hat{P}_{P}\right)\right\}^{2}+\left\{w_{3}\left(T_{P_{50}}-\hat{T}_{P_{50}} \right)\right\}^{2}+\left\{w_{4}\left(T_{P_{10}}-\hat{T}_{P_{10}} \right)\right\}^{2}
    \end{equation}

    \begin{equation*}
        \text{where,}~X=\left\{k_{1}, c_{1}, w_{11}, w_{12}, d_{1}, h_{1}, g_{1}, K_{AT_{De}}\right\}.
    \end{equation*}

    Here, $f(X)$= Objective function, $X$= Set of decision variables. $\hat{T}$= Time period of the simulated P waves, $\hat{P}_{P}$= Peak height of the simulated P waves, $\hat{T}_{P_{50}}$= Width at the 50\% height of the simulated P waves, and $\hat{T}_{P_{10}}$= Width at the 10\% height of the simulated P waves. Here, $w_{1}, w_{2}, w_{3}, \text{ and } w_{4}$ represent the weights of the different components in the objective function. These weights are introduced to maintain equal scaling between all components. Also, they are adjusted according to the importance of the individual components. 
    
    \item Phase 3: In Phase 3, the FHN equation corresponding to the QRS complex is optimized. Here the stimulus $I(t)$ is generated from the positive peak of the velocity component of the VDP equation for the HP node. This VDP equation has a coupling with the VDP equation for the AV node. In addition, the VDP equation corresponding to the AV node has a coupling with the SA node. Therefore, the optimized values of the SA node from Phase 1 and predefined values of AV and HP node VDP equations (except coupling constants and delays) are used to generate the velocity component of the VDP equation for the SA node. For this phase, we have assumed that coupling constants, $K_{\scriptscriptstyle SA-AV}= K_{\scriptscriptstyle AV-HP} \approx f_{1}$, and delays $\tau_{\scriptscriptstyle SA-AV}=\tau_{\scriptscriptstyle AV-HP}=0$. 
    \\~\\
    The most important features of the QRS complex in an ECG signal are (1) Average Peak height of the QRS waves ($P_{R}$), (2) Average width of the QRS waves at 50\% height of the peaks ($T_{R_{50}}$), (3) Average width of the QRS waves at 10\% height of the peaks ($T_{R_{10}}$), and (4) Average time period ($T$).  Thus the array by which the FHN equation for the QRS complex is optimized is $\left\{T, P_{R}, T_{R_{50}}, T_{R_{10}}, \right\}$. Thus the objective function for Phase 3 can be written as,


    \begin{equation}
        \text{Minimize,}~f(X)=\left\{w_{1}\left(T-\hat{T} \right)\right\}^{2}+\left\{w_{2}\left(P_{R}-\hat{P}_{R}\right)\right\}^{2}+\left\{w_{3}\left(T_{R_{50}}-\hat{T}_{R_{50}} \right)\right\}^{2}+\left\{w_{4}\left(T_{R_{10}}-\hat{T}_{R_{10}} \right)\right\}^{2} 
    \end{equation}

    \begin{equation*}
        \text{where,}~X=\left\{k_{3}, c_{3}, w_{31}, w_{32}, b_{3}, d_{3}, h_{3}, g_{3},K_{VN_{De}}\right\}.
    \end{equation*}

    Here, $\hat{T}$= Time period of the simulated QRS waves, $\hat{P}_{R}$= Peak height of the simulated QRS waves, $\hat{T}_{R_{50}}$= Width at the 50\% height of the simulated QRS waves, $\hat{T}_{R_{10}}$= Width at the 10\% height of the simulated QRS waves.
    Here, $w_{1}, w_{2}, w_{3} \text{ and } w_{4}$ represent the weights of the different components in the objective function. These weights are introduced to maintain equal scaling between all components. Also, they are adjusted according to the importance of the individual components. 
    
    \item Phase 4: FHN equation for T waves is optimized in this phase. Here also, the stimulus ($I(t)$) of this FHN equation depends on the velocity component of the VDP equation for the HP node. Similarly to Phase 3, the velocity component of the HP is estimated through the optimized parameter values of the SA node and prefixed values of the AV and HP nodes. In addition, the coupling constants and delays are assumed as $K_{\scriptscriptstyle SA-AV}= K_{\scriptscriptstyle AV-HP} \approx f_{1}$ and $\tau_{\scriptscriptstyle SA-AV}=\tau_{\scriptscriptstyle AV-HP}=0$. 

    As previously, the main four optimizing criteria are (1) Average peak height of the T waves ($P_{T}$), (2) Average width of the T waves at 50\% height of the peaks ($T_{T_{50}}$), (3) Average width of the T waves at 10\% height of the peaks ($T_{T_{10}}$), and (4) Average time period ($T$). We have observed that most of the T waves in the ECG signals are left skewed. Thus we have considered the average time interval between the position of the peak and the position where the height is 50\% of the peak of the T wave to its left ($T_{T_{sk}}$) to include this feature. The array for the optimization can be written as, $\left\{T, P_{T}, T_{T_{50}}, T_{T_{10}}, T_{T_{sk}} \right\}$. The objective function for Phase 4 can be written as, 

    \begin{equation}
        \begin{split}
        \text{Minimize,}~f(X)=\left\{w_{1}\left(T-\hat{T} \right)\right\}^{2}+\left\{w_{2}\left(P_{T}-\hat{P}_{T}\right)\right\}^{2}+\left\{w_{3}\left(T_{T_{50}}-\hat{T}_{T_{50}} \right)\right\}^{2}+\left\{w_{4}\left(T_{T_{10}}-\hat{T}_{T_{10}} \right)\right\}^{2}+\\ \left\{w_{5}\left(T_{T_{sk}}-\hat{T}_{T_{sk}} \right)\right\}^{2}
        \end{split}
    \end{equation}

    \begin{equation*}
        \text{where,}~X=\left\{k_{4}, c_{4}, w_{41}, w_{42}, d_{4}, h_{4}, g_{4}, K_{VN_{Re}}\right\}.
    \end{equation*}

    $\hat{T}$= Time period of the simulated T waves, $\hat{P}_{T}$= Peak height of the simulated T waves, $\hat{T}_{T_{50}}$= Width at the 50\% height of the simulated T waves, $\hat{T}_{T_{10}}$= Width at the 10\% height of the simulated T waves, and $\hat{T}_{T_{sk}}$= average time interval between the position of the peak and the position where the height is 50\% of the peak of the T wave to its left. Here, $w_{1}, w_{2}, w_{3}, w_{4} \text{ and } w_{5}$ represent the weights of the different components in the objective function. These weights are introduced to maintain equal scaling between all components. Also, they are adjusted according to the importance of the individual components.

    \item Phase 5: In this phase, the delays have been tuned according to the ECG data. Here we have assumed that $\tau_{\scriptscriptstyle SA-AV}=\tau_{\scriptscriptstyle AV-HP}=\tau$. Thus this system has two delays, $\tau$ and $\tau_{T}$ for optimization. Before optimization, we have generated the whole ECG signal using the optimized values of the parameters from the previous phases and prefixed parameter values for the VDP equations of AV and HP nodes. Also, the iso-electric baseline has been fixed at zero and the delays are initialized with zeros. The array of features by which the delays are optimized is $\left\{T_{\scriptscriptstyle PR}, T_{\scriptscriptstyle RT}\right\}$. Thus the objective function can be written as,

    \begin{equation}
        \text{Minimize,}~f(x)=\left\{w_{1}\left(T_{PR}-\hat{T}_{PR} \right)\right\}^{2}+\left\{w_{2}\left(T_{RT}-\hat{T}_{RT}\right)\right\}^{2}
    \end{equation}
\noindent

\begin{equation*}
    \text{where},~X=\left\{\tau, \tau_{T}\right\}.
\end{equation*}

Here, $\hat{T}_{PR}$= Peak to peak distance between the P wave and QRS complex of the simulated data and $\hat{T}_{RT}$= Peak to peak distance between the QRS complex and T wave of the simulated data. Also, $w_{1} \text{ and } w_{2}$ represent the weights of the objective function.

    \item ECG generation: In this phase, the final ECG is generated. The action potentials of the pacemaker cells are generated by the VDP equations of SA, AV, and HP nodes. For the VDP equation of the SA node, the optimized parameter values are used from Phase 1. For the VDP equations of the AV and HP nodes, prefixed values of the parameters are used. Further, the coupling constants of AV and HP nodes are assumed to be equal to the frequency of the SA node ($K_{\scriptscriptstyle SA-AV}= K_{\scriptscriptstyle AV-HP} \approx f_{1}$), and the delays are estimated from Phase 5. 
    
    From the velocity component of the SA node, P waves are generated by the corresponding FHN equation and optimized parameter values from Phase 2. Also, QRS and T waves are generated from their respective FHN equations (with optimized parameter values from Phase 3 and Phase 4) using the velocity component of the HP node as the stimulus. 

    In an ECG, the isoelectric baseline may not be fixed and can vary with time. However, to maintain simplicity we have assumed a constant isoelectric baseline does not vary with time. The variation of the isoelectric baseline ($B\ell$) has been chosen according to the data. Thus the final ECG has been generated by the superposition of all waves and with the inclusion of a proper baseline variation. 

    \begin{equation}
        \text{ECG}=B\ell+\text{P wave}+\text{QRS complex}+\text{T wave}.
    \end{equation}
\end{itemize}

\subsection{Genetic Algorithm (GA)}

In this subsection, we have briefly discussed the Genetic Algorithm (GA) which has been implemented to optimize the parameters of all different phases that have been discussed earlier. GA is a well-known optimization algorithm. Here, the objective function, $f(X)$ has been optimized with respect to the decision variables $X=\left\{x_{\scriptscriptstyle 1},x_{\scriptscriptstyle 2},\ldots x_{\scriptscriptstyle N}\right\}$ by GA. For this model, any decision variable $x_{\scriptscriptstyle i}$, belongs to the set of positive real numbers ($x_{\scriptscriptstyle i}\in \mathbb{R}|x_{\scriptscriptstyle i}\geq 0$) for all phases. All the variables are continuous variables. Thus we have judiciously chosen the upper and lower range of the values of all decision variables in our model.
\\~\\
In GA, the decision variable ($x_{\scriptscriptstyle i}$), set of decision variables ($X$), and set of solutions ($\left\{X_{1}, X_{2},\ldots, X_{N}\right\}$) are referred as gene, chromosome, and population respectively. We first set an initial population of size $N$ for GA by considering the random values of the decision variables from their respective ranges. The GA then proceeds with the actions listed below. 

\begin{enumerate}
    \item Parent selection: In this step, $R$ number of chromosomes are selected as parents from the population according to their fitness values 
    and the rest $N-R$ number of chromosomes are discarded. There are many methods to select parents from the population. Here, we have used the `Tournament Selection' method. In this selection technique, $Y$ number of solutions are randomly chosen from the population for the tournament. Then the best solution between the $Y$ number of solutions is selected as a parent solution and then excluded from the population set. By repeating this process, the set of parent population of size $R$ is obtained.

    \item Generation of children: The new solutions that are generated from the parent population are called the children. The new solutions can be generated by the crossover technique. Here we have used the real coded crossover technique with single-point crossover. Using this method, $N-R$ number of children are generated.

    \item Mutation: We can generate new solutions by crossover technique. However, this method only shuffles the pre-existing information in the parent solutions and does not include any new information. Thus for the evolution of the solutions `Mutation' is necessary. In this step, every gene of a new solution is selected with a probability $P_{\scriptscriptstyle M}$ and some small value is added which is drawn from a normal distribution ($\mathcal{N}\left(0,\sigma_{i}^2\right), i=\text{index of the chosen gene}$).  

    \item New population: The $R$ number of parents and $N-R$ number of children (after mutation) are now considered as the new population.
\end{enumerate}

The steps 1 to 4 are repeated until the algorithm meets some pre-defined stopping criteria. We have implemented GA for each phase that was discussed earlier and the best values of the parameters were chosen. Also, the parameter values for GA like $N, R, \text{and } P_{M}$ are chosen and tuned properly for each phase.
\section{Results and Discussions}\label{results}

In this section, we have discussed the results obtained by fitting the data with genetic algorithm (GA) optimization process. We had already discussed the Ryzhii model and the modifications introduced to it in Sec. \pref{Formulation}. The details of the GA had also been discussed in Sec. \pref{Mod_fit}. We have presented the fits that we have obtained with our optimization process in this section. The data representing normal sinus rhythm is discussed first with a representative fit, following which we have tested data exhibiting sinus tachycardia and sinus bradycardia.  These are still classified as sinus rhythm but with heart rates lying outside the normal sinus interval range of 60-100 bpm in adults \cite{kusumoto2020ecg,wagner2001marriott}. We have taken data from the MIT-BIH Arrhythmia database \cite{PhysioNet} which exhibits slight deviations from the normal sinus rhythm to test how far we can stretch the capabilities of the fitting algorithm. We have also excluded the arrhythmia portions of the data that we are working with as our algorithm does not support heart rate variability. Finally, our algorithm has also been tested with pathological data belonging to the cases of $1^{\text{st}}$ Degree AV Block and $3^{\text{rd}}$ Degree AV Block. \\~\\ 

\subsubsection*{Normal Sinus Rhythm}\noindent We start with the normal sinus rhythm data-fits where we have used data from the MIT-BIH Normal Sinus Rhythm Database \cite{PhysioNet}. As seen from Fig.~\pref{fig:NSR_datafit} given below, we can conclude that the SA node sets up the heart rate and conducts the impulses to the AV and HP nodes, as a result of which there is an orderly fashion of P-Wave followed by the QRS-complex and T-Wave. This data-set has a heart rate lying between 60-100 bpm ($\sim 95$ bpm) \cite{kusumoto2020ecg,litfl,wagner2001marriott} and hence falls under the normal sinus rhythm. We also note that our algorithm does not support frequency variation but instead takes average representative frequencies of the constituent waves.
\begin{figure}[H]
    \centering
    \includegraphics[scale=0.35]{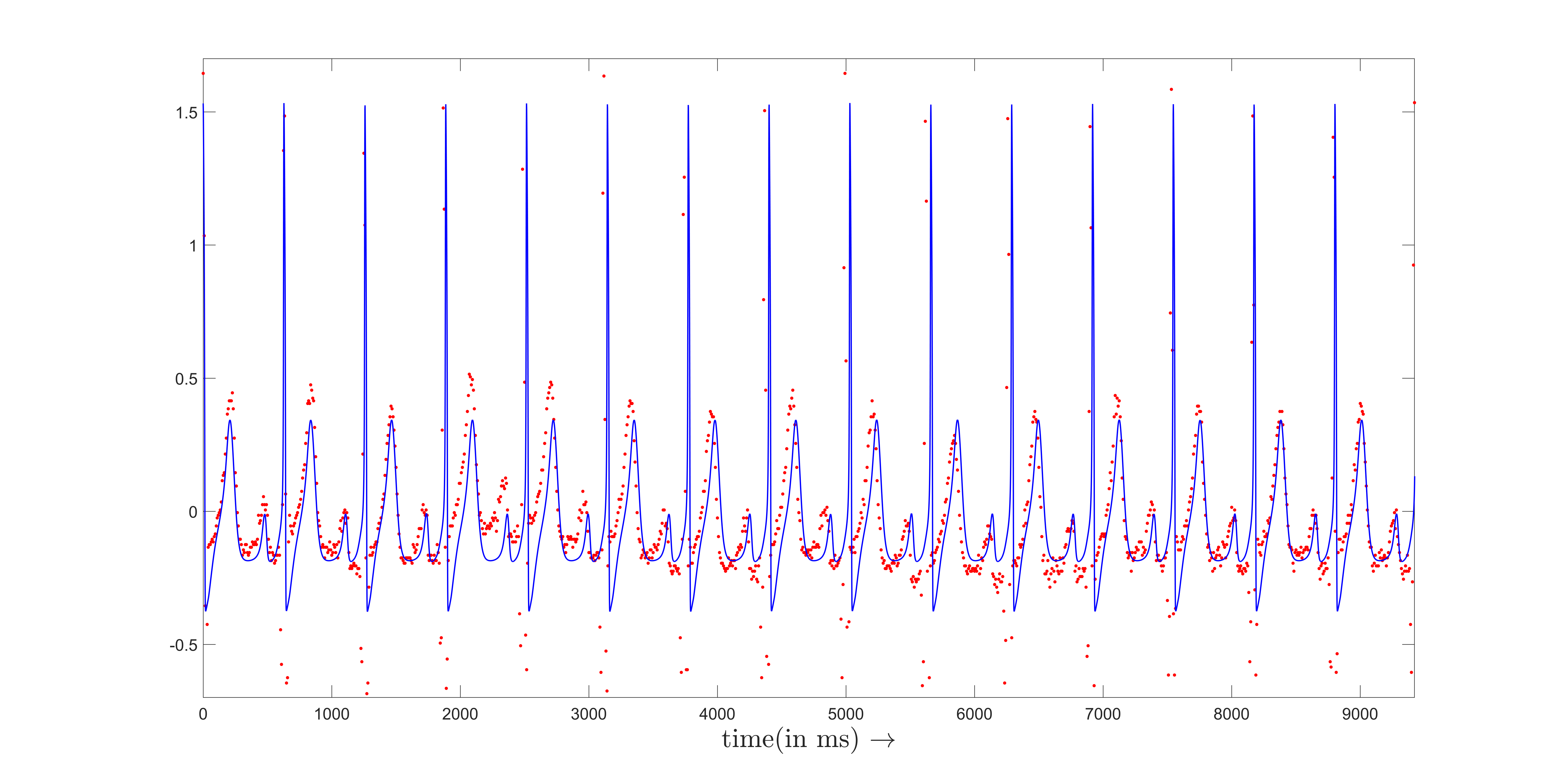}
    \caption{Data Fitting of Normal Sinus Rhythm }
    \label{fig:NSR_datafit}
\end{figure}
\noindent
\subsubsection*{Sinus Tachycardia}
\begin{figure}[H]
    \centering
    \includegraphics[scale=0.35]{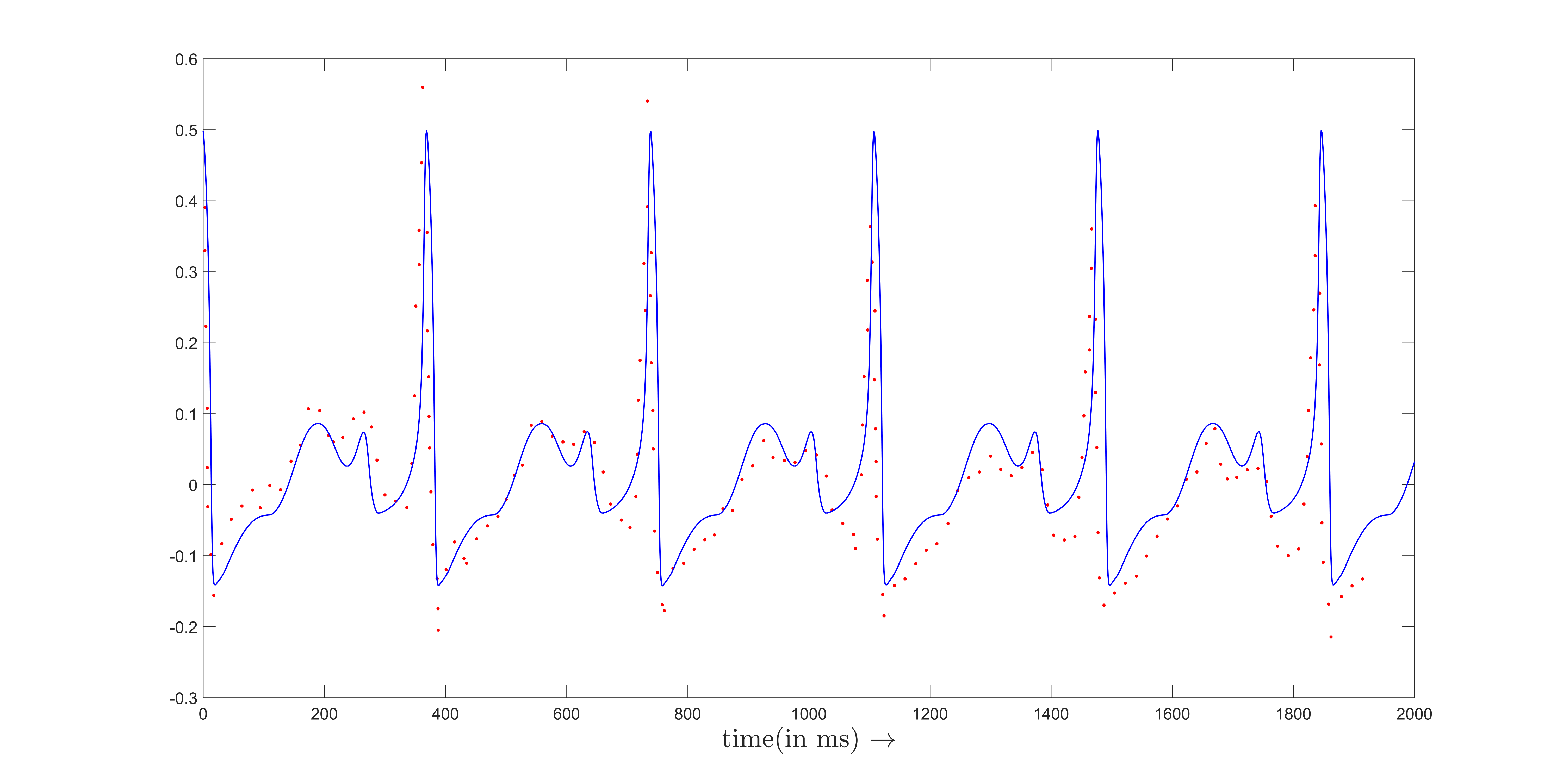}
    \caption{Data Fitting of Sinus Tachycardia }
    \label{fig:Tachy_datafit}
\end{figure}

\noindent The next case that we discuss is that of sinus tachycardia. This condition is characterized by heart rates above 100 bpm in adults \cite{kusumoto2020ecg,litfl,wagner2001marriott}. The data is taken from \cite{litfl} and it shows a heart rate of $\sim 160$ bpm and still normal sinus ordering of constituent waves. Since the SA node follows such a high frequency, it results in the shortening of the time period of the action potential. As a result of this, there is a decrease in the PR and RT intervals. This can also be seen in the closing in of the P and T waves in Fig.~\pref{fig:Tachy_datafit}. 
\noindent
\subsubsection*{Sinus Bradycardia}\noindent Following tachycardia, we have studied the slow end of normal sinus rhythm i.e. sinus bradycardia. This condition is characterized by heart rates below 60 bpm in adults \cite{kusumoto2020ecg,litfl,wagner2001marriott}. The data \cite{litfl} shown exhibits a heart rate of $\sim 59$ bpm and yet shows normal sinus ordering of constituent waves. Since the SA node follows such a low frequency, it results in lengthening of the time period of the action potential. As a result of which there is an increase in the PR and RT intervals. 
 \begin{figure}[H]
    \centering
    \includegraphics[scale=0.35]{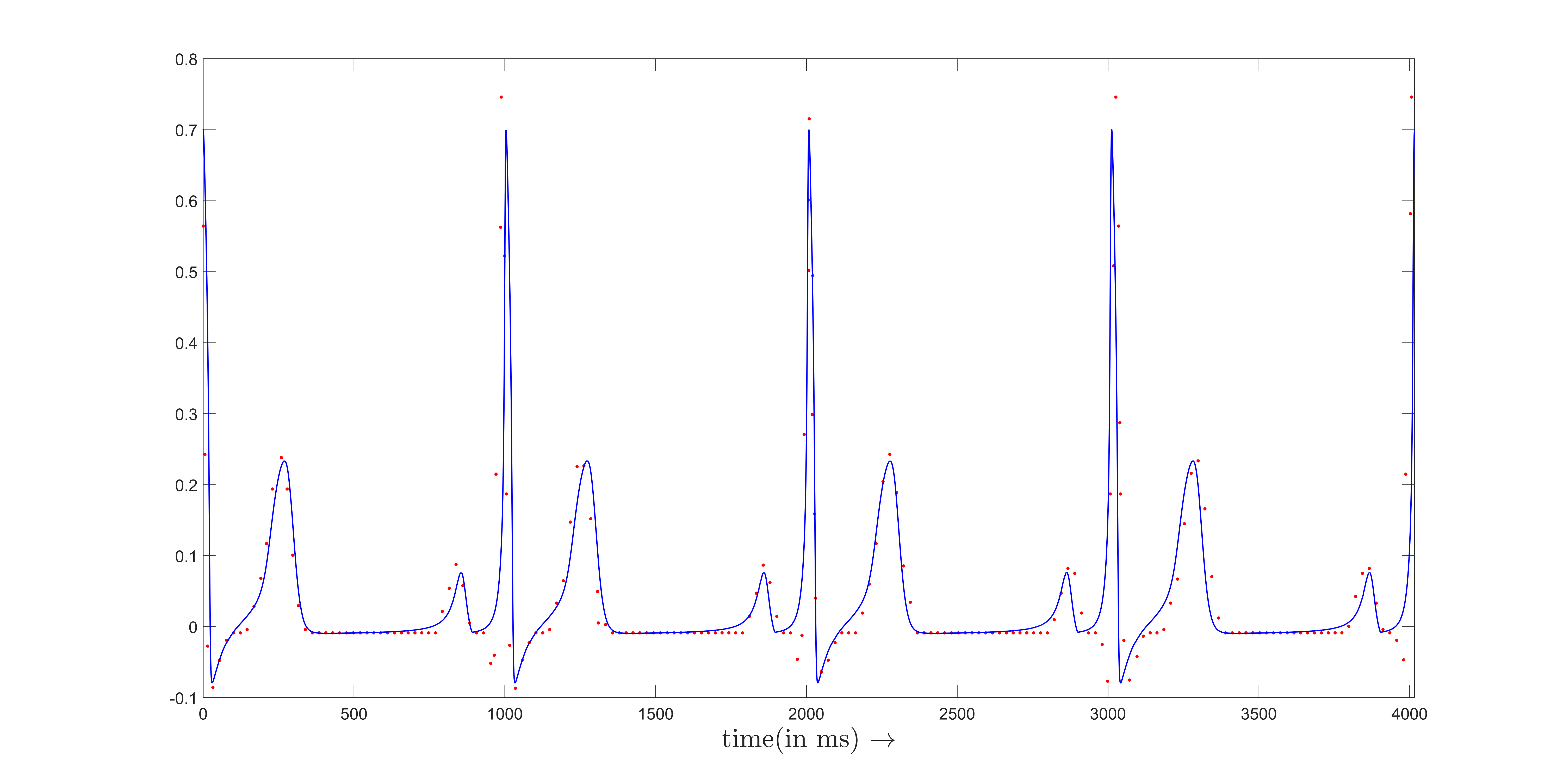}
    \caption{Data Fitting of Sinus Bradycardia }
    \label{fig:Brady_datafit}
\end{figure}
 \subsubsection*{$1^{\text{st}}$ degree AV Block}\noindent Next we discuss fits of pathological datasets. The dataset that we have examined first is of $1^{\text{st}}$ degree AV Block, which deals with some loss in synchronization between the pacemaker complexes. Hence there is some communication loss between the atria and the ventricles. This condition is characterized by PR intervals in the ECG above 200 ms \cite{litfl}. The normal sinus range of the PR interval is 120-200 ms \cite{kusumoto2020ecg,litfl,wagner2001marriott}. In case of the chosen data \cite{litfl}, the PR interval is $\sim 293$ ms. Such cases are generally asymptomatic except in cases where the PR interval exceeds 300 ms \cite{litfl}. These kinds of AV blocks are called `marked AV Block' \cite{litfl}. The longer PR interval results because of a delay existing in the conduction of the impulse from the SA node to the AV node. This situation seems similar to sinus tachycardia because of the closing in of P and T waves. However, the physiological difference is due to the former having a simply higher conduction rate and the latter is due to an increase in the SA-AV delay. 
 \begin{figure}[H]
    \centering
    \includegraphics[scale=0.35]{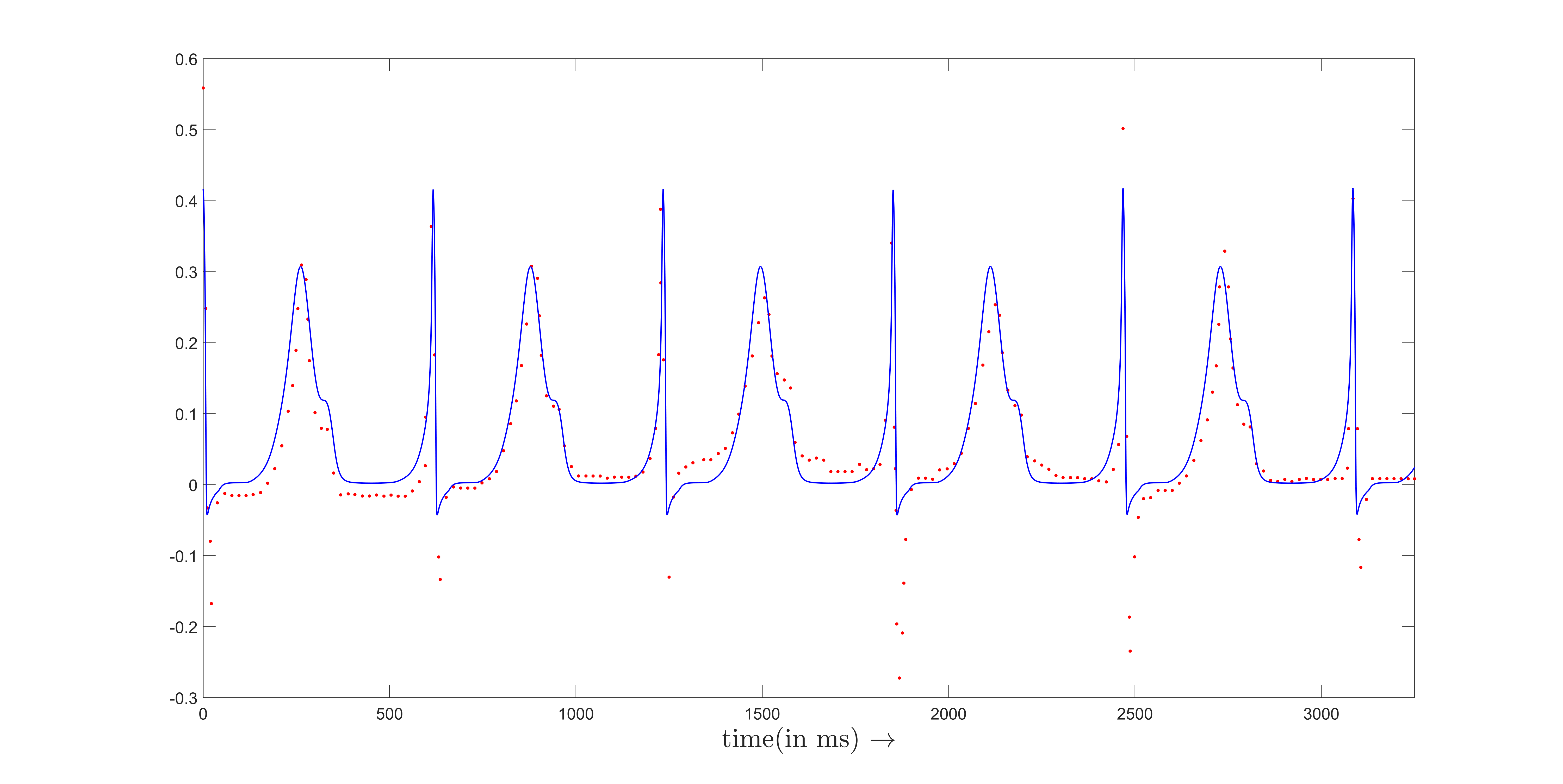}
    \caption{Data Fitting of $1^{\text{st}}$ Degree AV Block}
    \label{fig:1stdegAV_datafit}
\end{figure}

\subsubsection*{$3^{\text{rd}}$ degree AV Block}\noindent The second case is of $3^{\text{rd}}$ degree AV Block which is characterized by a complete loss in synchronization between the SA and the AV nodes, resulting in complete communication loss between the atria and the ventricles \cite{kusumoto2020ecg,litfl,wagner2001marriott}. This results in independent frequencies of atria and ventricular wave components. From the data  that we have analyzed \cite{litfl}, we can see that the P waves are driven by the SA node intrinsic frequency and the QRS and T waves are driven by the HP node intrinsic frequency. Thus to replicate this feature we set the coupling constants, $K_{SA-AV}$ and $K_{AV-HP}$ to zero ($K_{SA-AV}=K_{AV-HP}=0$) in the respective pacemaker blocks.
 \begin{figure}[H]
    \centering
    \includegraphics[scale=0.35]{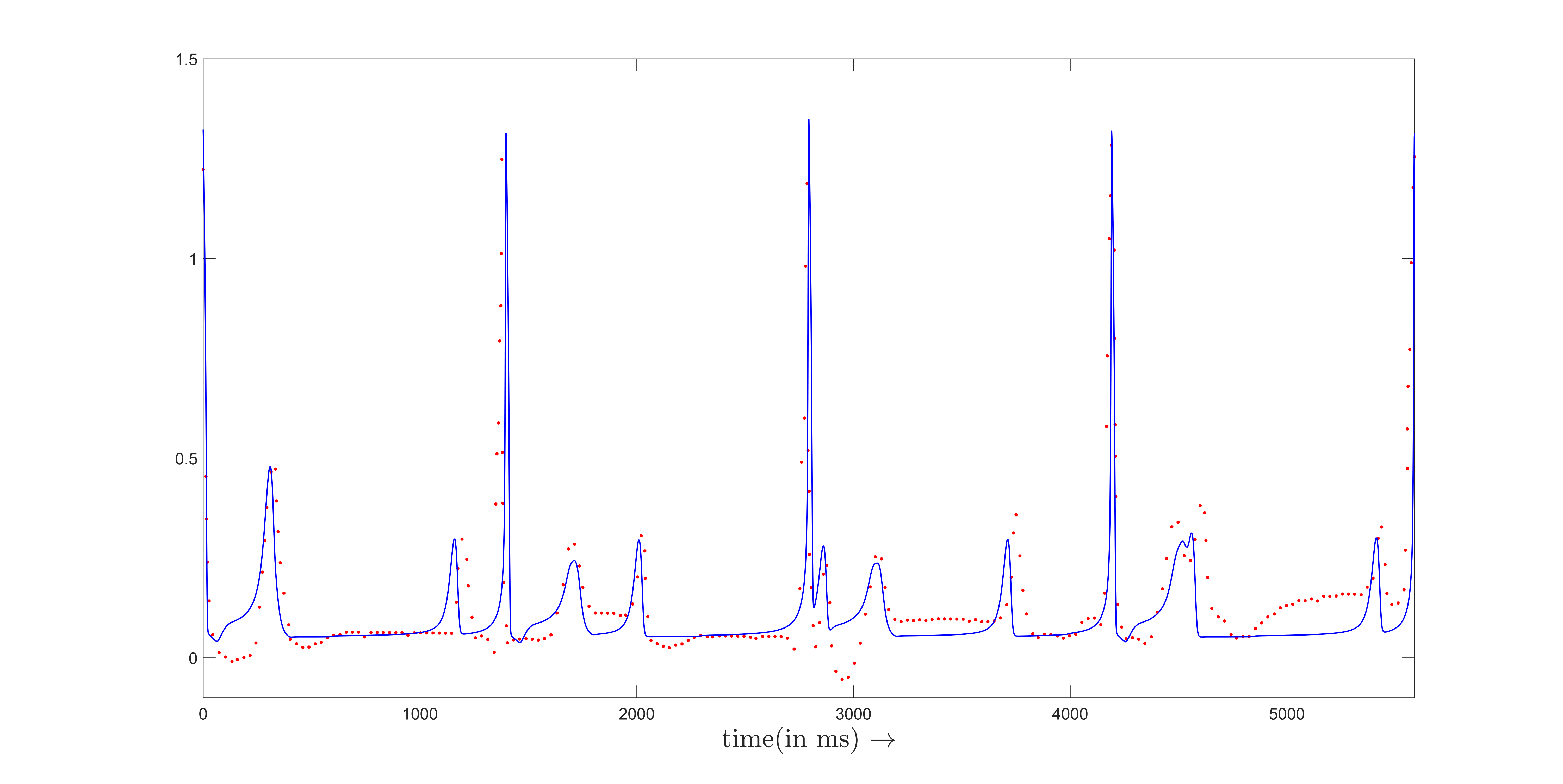}
    \caption{Data Fitting of $3^{\text{rd}}$ Degree AV Block}
    \label{fig:3rddegAV_datafit}
\end{figure}

\subsubsection*{Test Data}
\begin{figure}[H]
    \begin{subfigure}{.5\textwidth}
        \centering
        \includegraphics[scale=0.6]{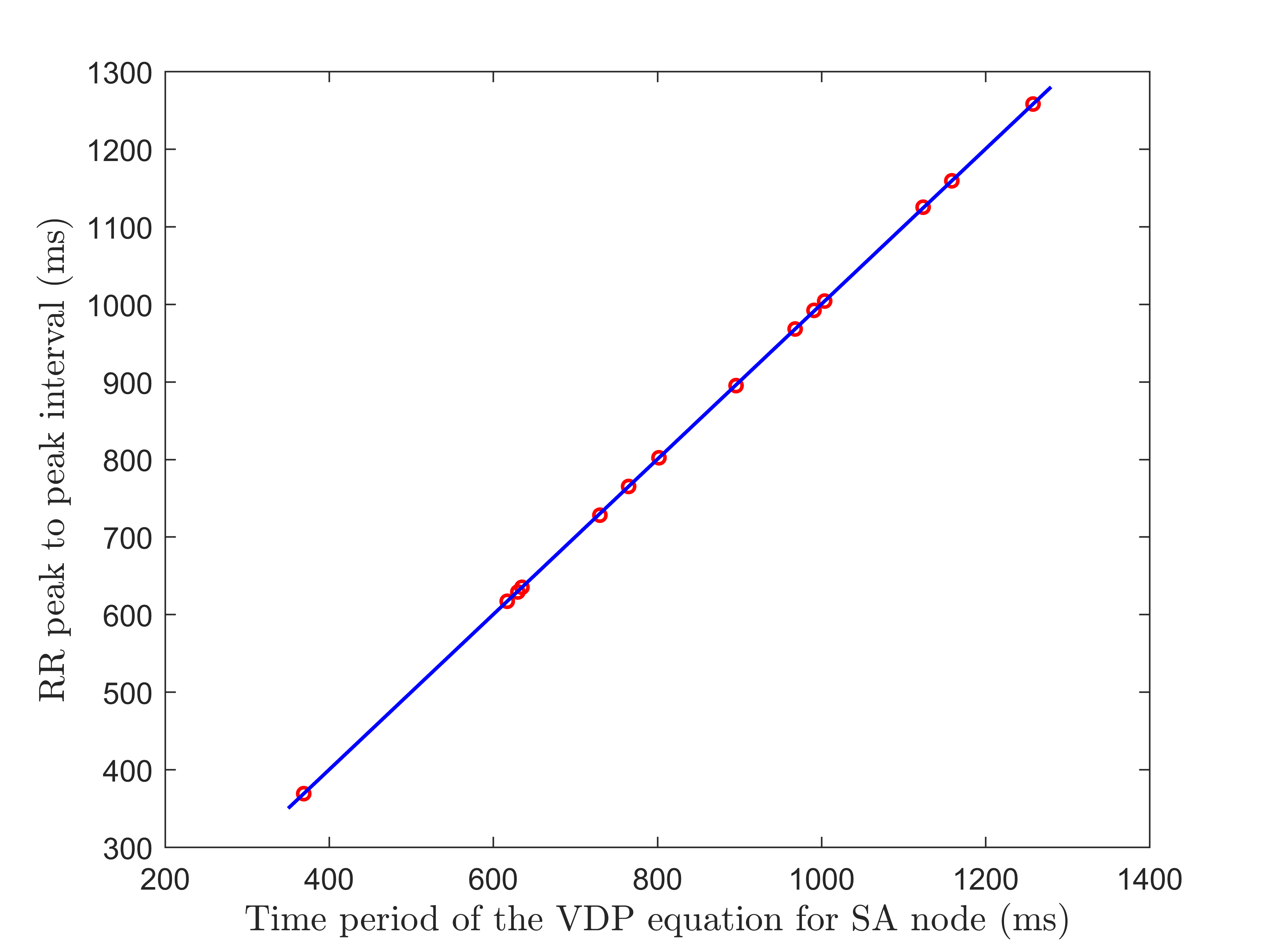}
        \caption{}
        \label{RR_freq}
    \end{subfigure}
    \begin{subfigure}{.5\textwidth}
        \centering
        \includegraphics[scale=0.6]{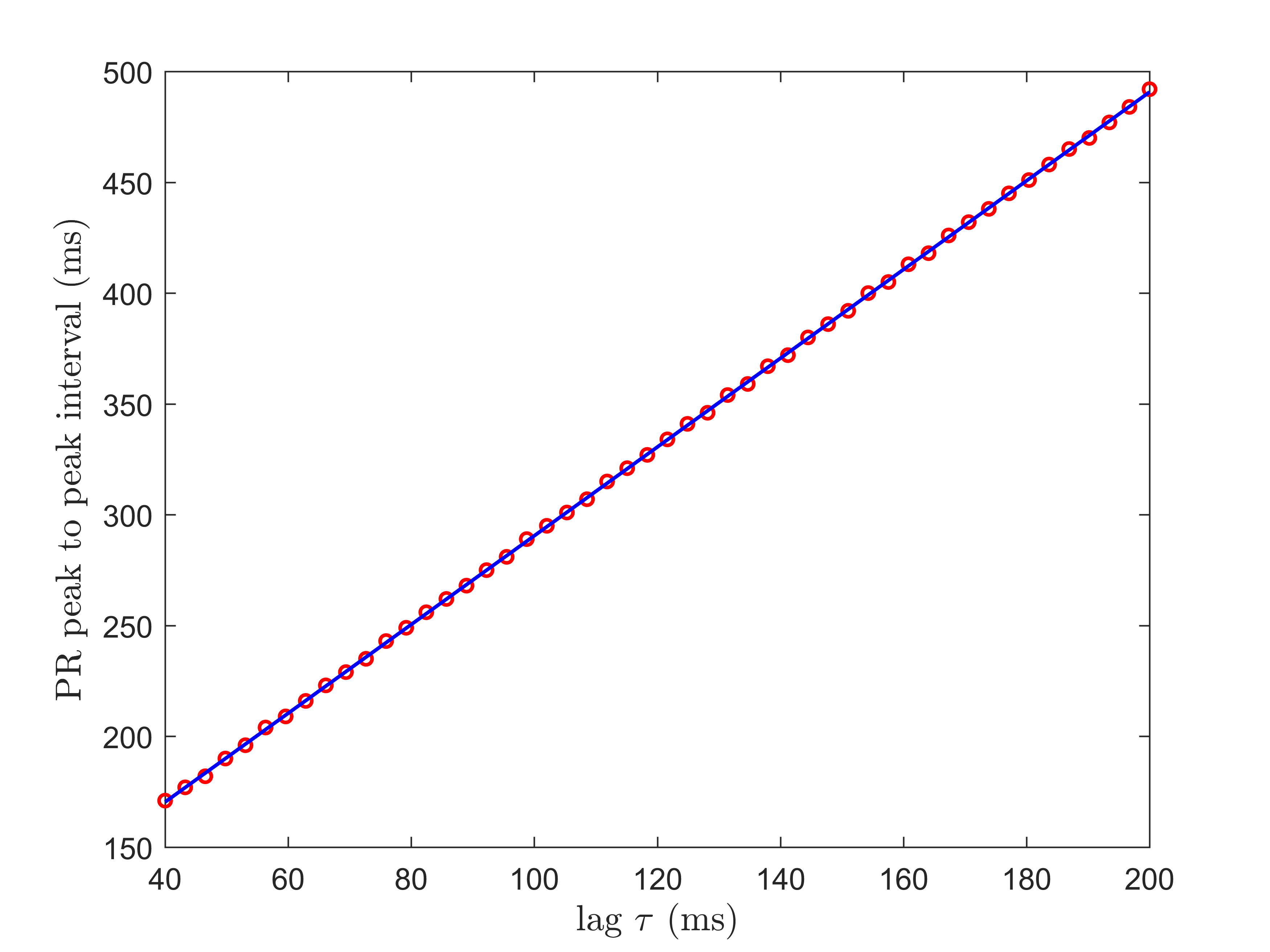}
        \caption{}
        \label{PR_lag}
    \end{subfigure}
    \caption{(a) Graph of RR peak-to-peak interval and time period of the VDP equation for the SA node for different ECG fits which is represented with red dots. The solid blue line represents the linear fit. (b) Variation of peak-to-peak PR-Interval with lag $\tau$ for a sinus rhythm which is represented with red dots. The solid blue line represents the linear fit. The gradient ($m$) and intercept ($c$) of this straight line are 2 and 0.09 respectively.}
    \label{fig:analysis}
\end{figure}

\noindent These two datasets were taken from the MIT-BIH arrhythmia database \cite{PhysioNet} to check how well our genetic algorithm based framework handles ECG data which is not classified. The first dataset has an RR interval of 1125 ms (heart rate of $\sim$53 bpm) and PR interval of 139 ms and RT interval of 852 ms. The second dataset has an RR interval of 635 ms (heart rate of $\sim$94 bpm) and PR interval of 173 ms and RT interval of 222 ms. 
 \begin{figure}[H]
    \centering
    \includegraphics[scale=0.35]{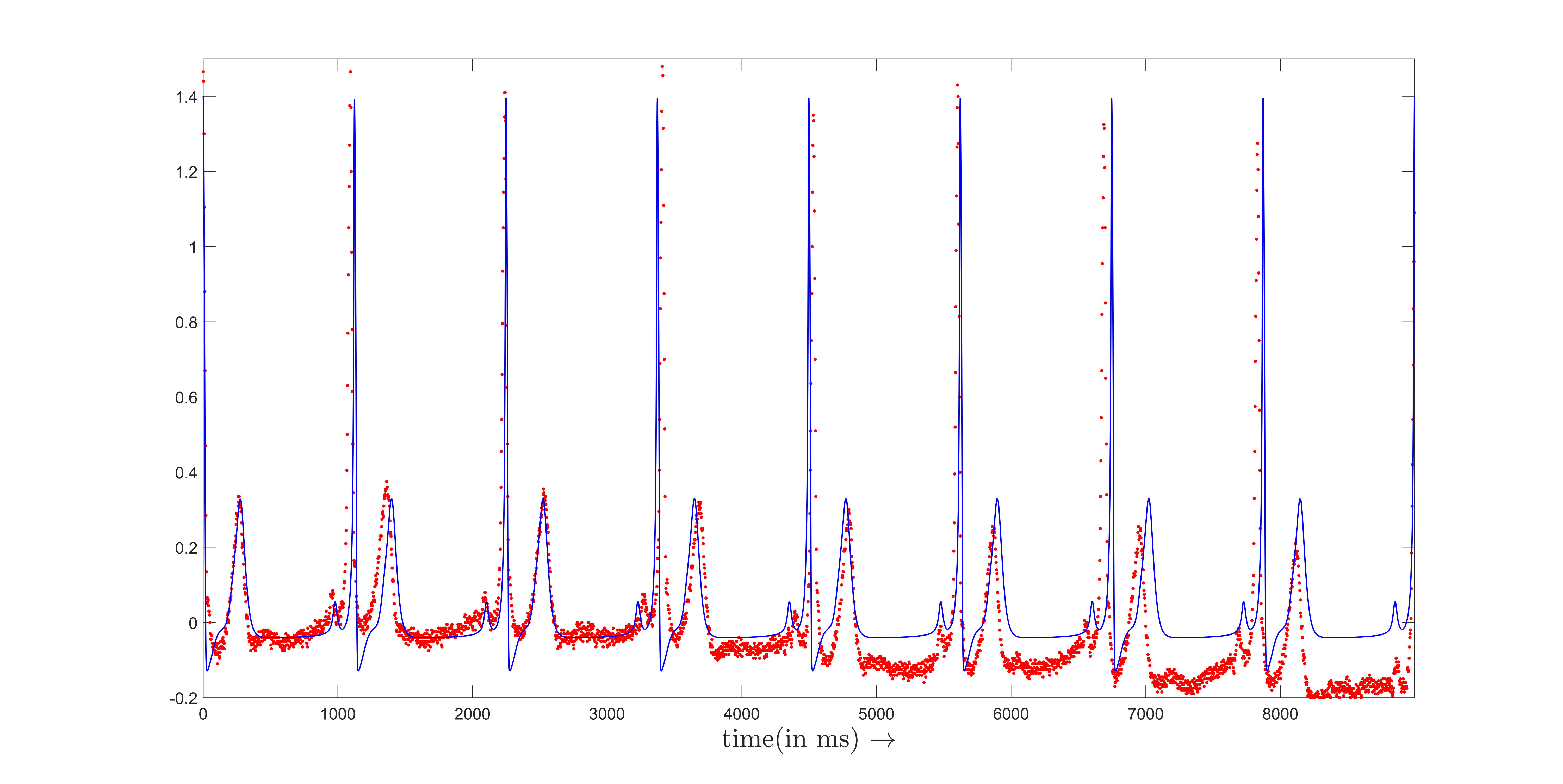}
    \caption{Test Data (1), taken from MIT-BIH Arrhythmia database \cite{PhysioNet} and fitted with our model using GA. Red dots represent ECG data, solid blue line represents model fit to the data}
    \label{fig:R1_datafit}
\end{figure}
\begin{figure}[H]
    \centering
    \includegraphics[scale=0.35]{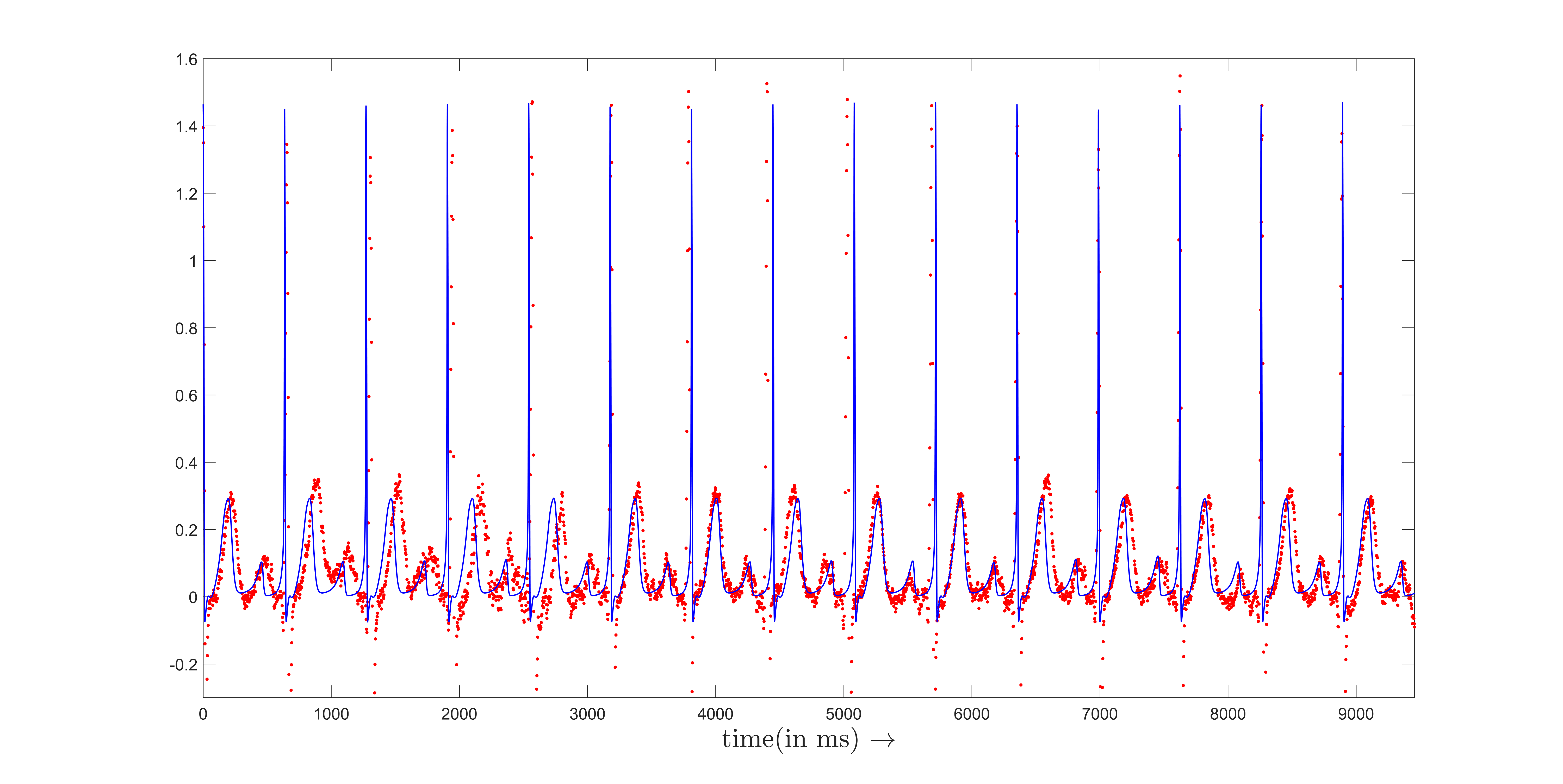}
    \caption{Test Data (2), taken from MIT-BIH Arrhythmia database \cite{PhysioNet} and fitted with our model using GA. Red dots represent ECG data, solid blue line represents model fit to the data}
    \label{fig:R2_datafit}
\end{figure}
\noindent The two heart conditions, tachycardia and bradycardia can be classified using the frequency of an ECG signal. As mentioned earlier, if the frequency of an ECG is less than 60 bpm or greater than 100 bpm, the heart belongs to the regime of tachycardia or bradycardia respectively \cite{kusumoto2020ecg,litfl,wagner2001marriott}. It is assumed earlier that the frequency of the synthetic ECG of this model is completely controlled by the frequency of the VDP equation for the SA node (Eq.~\pref{eq:SA_eqn}). The relationship between the peak-to-peak RR interval and time period of the VDP equation for the SA node is linear with gradient and intercept as one and zero respectively, as shown in Fig.~\pref{RR_freq}. This result agrees with our assumption. The two parameters of this equation which mainly control the frequency are $a_{1}$ and $f_{1}$. However other parameters of this equation also affect the frequency. Thus the frequency of the VDP equation of the SA node is the most important factor which influences the generation of the synthetic ECG signal with a normal heart rate.  

\noindent In this model, the PR interval mainly depends on two lags, $\tau_{\scriptscriptstyle SA-AV}$ and $\tau_{\scriptscriptstyle AV-HP}$. While fitting our model to the given data sets, it was assumed that $\tau_{\scriptscriptstyle SA-AV}=\tau_{\scriptscriptstyle AV-HP}=\tau$ (as shown in Sec.~\pref{Mod_fit}). In addition, it is also noted that the PR interval is a function of $\tau_{\scriptscriptstyle SA-AV}+\tau_{\scriptscriptstyle AV-HP}$ (refer to Fig.~\pref{delay_PR}), which means that it is dependent on $2\tau$. Fig.~\pref{PR_lag} shows the linear increment of the peak-to-peak PR interval with $\tau$. To generate this figure, we have taken the fitted parameter values for the normal sinus rhythm which are shown in Fig.~\pref{fig:NSR_datafit}. For all of these fitted cases, the peak-to-peak PR interval increases linearly with $\tau$ with a gradient, $m\simeq 2$. However, the values of the intercepts ($c$) change from case to case and the main difference between these linear relationships for different fits are through these values of $c$. Hence, the effect of all the other parameters of this model (except $\tau$) on the PR interval are inherited by $c$. For the classification of the $1^{\text{st}}$ degree AV block from the normal sinus rhythm, the PR interval is the prime indicator is usually longer compared to the normal sinus rhythm \cite{kusumoto2020ecg,litfl,wagner2001marriott}. To generate the synthetic data for the $1^{\text{st}}$ degree AV block, the lag $\tau$ should be tuned according to the linear relationship between PR interval and $\tau$ for that particular set of parameters.

\section{Conclusion} \label{conclusion}

In this section, we have briefly discussed the aim of this paper, the system to be modeled and the algorithm we have used to fit the data with our proposed model. This was a pilot study that we undertook to see how this model ($\tau$-FHN model) , a modified form of the Ryzhii \cite{ryzhii2014heterogeneous} could reproduce  real ECG signals of the human heart within a certain range of bpm. The main features, importance and future scope of our model are discussed. Finally, the limitations and further improvement possibilities are also mentioned.  Other DDE-based models have successfully reproduced the ECG waveforms but lack the proper fitting in the real ECG data. The primary aim of this work is to obtain the best fit of ECG waveforms for different heart conditions using genetic algorithm (GA) to understand the behaviour of the model ($\tau$-FHN model) equations and their parameters from the perspective of real ECG waveforms. 
\\~\\
\noindent In this work, the cardiac conduction system is modeled by separating the heart into two different components. The first component is the pacemaker system which acts as a driver of the heart. The pacemaker system comprises of three nodes, SA, AV and HP which generate action potentials for the heart. The second component is the non-pacemaker cells of atria and ventricles which depolarize and repolarize under the effect of the action potentials. The electrical activity of these non-pacemaker cells gets recorded as the ECG wave components. The model of the action potentials generated from the pacemakers and the depolarization and repolarization of the non-pacemaker cells of the atria and ventricles are considered from the model based on the work by Ryzhii \textit{et. al.} \cite{ryzhii2014heterogeneous}. The action potentials generated from the SA, AV and HP nodes are modeled with the VDP equations separately (Eq.~\pref{eq:SA_eqn}, \pref{eq:AV_eqn} and \pref{eq:HP_eqn}). The depolarization and the repolarization of the non-pacemaker cells of the atria and ventricles are modeled with the four sets of the FHN equations (Eq.~\pref{Eqn:P_Wave}, \pref{Eqn:Ta_Wave}, \pref{Eqn:QRS_Wave} and \pref{Eqn:T_Wave}). The SA node is the primary pacemaker of the heart and the other two pacemakers run with the same frequency as the SA node. Thus in this model, the VDP equations corresponding to the AV and HP nodes, are coupled with the VDP equations of the SA and AV nodes respectively with a delay term. 
\\~\\
\noindent Here we note that one of the drawbacks of the previous FHN model by Ryzhii \textit{et. al.} \cite{ryzhii2014heterogeneous} was that it sometimes failed to generate the QRS and T waves with the proper interval. Thus, to correct this feature, an essential modification to these equations ventricles (Eq.~\pref{Eqn:QRS_Wave} and \pref{Eqn:T_Wave}) was introduced in the form of the addition of a delay term. This delay mainly increases the RT interval of an ECG signal, as shown in  Fig.~\pref{fig:model_lags}.
\\~\\
\noindent Next we have  summarized the results obtained below.

\begin{itemize}
   
   \item In Fig.~\pref{delay_PR}, we have shown that the delays of the VDP equations of the AV and HP nodes ($\tau_{\scriptscriptstyle SA-AV}$ and $\tau_{\scriptscriptstyle AV-HP}$) mainly increase the PR interval and their effects on the PR interval are additive. 

    \item The delay $\tau_{T}$ increases the RT interval as shown in Fig.~\pref{fig:model_lags} in addition to an increase in the T wave peak as seen in the inset of Fig.~\pref{fig:model_lags}. However, this increment in T wave peak is very small and can be safely neglected.

    \item We have also developed a proper framework to fit this model with the ECG data for different cases using the genetic algorithm (GA). At first, we chose the essential parameters of our model and then divided our model into different parts for fitting (detailed discussion in Sec.~\pref{Mod_fit}). We have used this scheme to fit the ECG data of multiple heart conditions like normal sinus rhythm, sinus tachycardia, sinus bradycardia and first-degree AV block reasonably well, which is discussed im detail in Sec.~\pref{results}.
    
    \item The fited results of this model  show that the frequency of the generated ECG completely depends on the frequency of the VDP equation for the SA node (refer to Fig.~\pref{RR_freq}). Also, the PR interval is mainly dependent on the lag $\tau$ (refer to Fig.~\pref{PR_lag}). Thus the VDP equation for the SA node and the lag $\tau$ are the most important parts for generating a synthetic ECG signal.
    
    \item We have also shown that the model can generate synthetic data (using the proposed fitting framework) where the atrial and ventricular responses are not synchronized which can mimic the ECG of the third-degree AV block (Fig.~\pref{fig:3rddegAV_datafit}). 

    \item Finally, we have also fitted some unclassified ECG data which have been randomly chosen from the database \cite{PhysioNet}. Our model has shown reasonable agreement with the data in the form of good fits, as shown in Fig.~\pref{fig:R1_datafit} and Fig.~\pref{fig:R2_datafit} respectively.

\end{itemize}
\noindent A few other pathological conditions which falls under arrhythmia could not be fitted due to some specific reasons, in addition to loss of a set pattern in the ECG wave components due to arrhythmic beats. For example,  atrial fibrillation had an absence of P-waves and variable ventricular rate. In case of atrial flutter the isoelectric baseline is replaced by sawtooth like flutter waves which are not supported by the system of DDEs. Ventricular fibrillation on the other hand is characterised by no identifiable P, QRS and T waves and chaotic deflections of varying amplitudes which sometimes reaches 500 bpm. Paced rhythm was also impossible to replicate due to deviation from normal sinus type waveforms. Also, a wide variety of conditions (e.g., pericarditis, myocarditis etc.) are diagnosed by a combination of different ECG leads and since we are only concerned with one ECG lead (MLII), our choice of data was limited. 
\\~\\ 
\noindent The model can further be improved by incorporating a heart-rate variability function that might help mitigate the smaller degree arrhythmic beats which exist in any clinical ECG. This will enable us to replicate and hence fit atrial fibrillation and other normal sinus data which has presence of arrhythmic beats. 
\\~\\ 
\noindent  Another problem we have faced is an unavailability of classified long term ECG data belonging to a wide range of heart conditions. In the future, we would  wish to collaborate with medical fraternities and use their expertise to get further insight about the ECG and cardiovascular diseases through this model and beyond.

\section*{Author contributions}
S. C. has written the program codes and A. G. has done the analyses for this work. All of the authors have contributed to the development of this work and planning and preparing this manuscript. Also, all of the authors have thoroughly checked and reviewed this manuscript before submitting it.

\section*{Acknowledgement}
S. C. would like to acknowledge the financial support provided by the University Grant Commission (UGC), Government of India, in the form of `CSIR-UGC NET-JRF'. One of the authors S. R. C. would like to thank St. Xavier's College (Autonomous), Kolkata for providing financial assistance in the form of `Intramural Research Grant for R\&D Projects'. All of the authors would like to thank St. Xavier’s College (Autonomous), Kolkata for hosting this project.

\bibliographystyle{unsrt}
\bibliography{ref}

\begin{thebibliography}{10}

\bibitem{vaduganathan2022global}
Muthiah Vaduganathan, George~A Mensah, Justine~Varieur Turco, Valentin Fuster,
  and Gregory~A Roth.
\newblock The global burden of cardiovascular diseases and risk: a compass for
  future health, 2022.
\newblock DOI: 10.1016/j.jacc.2022.11.005.

\bibitem{tsao2023heart}
Connie~W Tsao, Aaron~W Aday, Zaid~I Almarzooq, Cheryl~AM Anderson, Pankaj
  Arora, Christy~L Avery, Carissa~M Baker-Smith, Andrea~Z Beaton, Amelia~K
  Boehme, Alfred~E Buxton, et~al.
\newblock Heart disease and stroke statistics—2023 update: a report from the
  american heart association.
\newblock {\em Circulation}, 147(8):e93--e621, 2023.
\newblock DOI: 10.1161/CIR.0000000000001123.

\bibitem{pineiro2023world}
Daniel~Jos{\'e} Pi{\~n}eiro, Jagat Narula, Borjana Pervan, and Lisa Hadeed.
\newblock World heart day 2023: Knowing your heart, 2023.
\newblock DOI: 10.4103/ijmr.ijmr\_1689\_23.

\bibitem{kalra2023burgeoning}
Ankur Kalra, Arun~Pulikkottil Jose, Poornima Prabhakaran, Ashish Kumar, Anurag
  Agrawal, Ambuj Roy, Balram Bhargava, Nikhil Tandon, and Dorairaj Prabhakaran.
\newblock The burgeoning cardiovascular disease epidemic in
  indians--perspectives on contextual factors and potential solutions.
\newblock {\em The Lancet Regional Health-Southeast Asia}, 2023.
\newblock DOI: 10.1016/j.lansea.2023.100156.

\bibitem{kumar2020cardiovascular}
A~Sreeniwas Kumar and Nakul Sinha.
\newblock Cardiovascular disease in india: a 360 degree overview, 2020.
\newblock DOI: 10.1016/j.mjafi.2019.12.005.

\bibitem{niederer2019short}
Steven~A Niederer, Kenneth~S Campbell, and Stuart~G Campbell.
\newblock A short history of the development of mathematical models of cardiac
  mechanics.
\newblock {\em Journal of molecular and cellular cardiology}, 127:11--19, 2019.
\newblock DOI: 10.1016/j.yjmcc.2018.11.015.

\bibitem{hill1938heat}
Archibald~Vivian Hill.
\newblock The heat of shortening and the dynamic constants of muscle.
\newblock {\em Proceedings of the Royal Society of London. Series B-Biological
  Sciences}, 126(843):136--195, 1938.
\newblock DOI: 10.1098/rspb.1938.0050.

\bibitem{huxley1957muscle}
Andrew~F Huxley.
\newblock Muscle structures and theories of contraction.
\newblock {\em Progr Biophys Chem}, 7:255--318, 1957.
\newblock DOI: 10.1016/S0096-4174(18)30128-8.

\bibitem{julian1969activation}
Fred~J Julian.
\newblock Activation in a skeletal muscle contraction model with a modification
  for insect fibrillar muscle.
\newblock {\em Biophysical Journal}, 9(4):547--570, 1969.
\newblock DOI: 10.1016/S0006-3495(69)86403-9.

\bibitem{saeki1980transient}
Y~Saeki, K~Sagawa, and H~Suga.
\newblock Transient tension responses of heart muscle in ba2+ contracture to
  step length changes.
\newblock {\em American Journal of Physiology-Heart and Circulatory
  Physiology}, 238(3):H340--H347, 1980.
\newblock DOI: 10.1152/ajpheart.1980.238.3.H340.

\bibitem{parmley1967series}
William~W Parmley and Edmund~H Sonnenblick.
\newblock Series elasticity in heart muscle: Its relation to contractile
  element velocity and proposed muscle models.
\newblock {\em Circulation Research}, 20(1):112--123, 1967.
\newblock DOI: 10.1161/01.RES.20.1.112.

\bibitem{fung1970mathematical}
Yuan-Cheng Fung.
\newblock Mathematical representation of the mechanical properties of the heart
  muscle.
\newblock {\em Journal of biomechanics}, 3(4):381--404, 1970.
\newblock DOI: 10.1016/0021-9290(70)90012-6.

\bibitem{arts1979model}
Theo Arts, Robert~S Reneman, and Peter~C Veenstra.
\newblock A model of the mechanics of the left ventricle.
\newblock {\em Annals of biomedical engineering}, 7:299--318, 1979.
\newblock DOI: 10.1007/BF02364118.

\bibitem{guccione1991passive}
Julius~M Guccione, Andrew~D McCulloch, and LK~Waldman.
\newblock Passive material properties of intact ventricular myocardium
  determined from a cylindrical model.
\newblock 1991.
\newblock DOI: 10.1115/1.2894084.

\bibitem{guccione1995finite}
Julius~M Guccione, Kevin~D Costa, and Andrew~D McCulloch.
\newblock Finite element stress analysis of left ventricular mechanics in the
  beating dog heart.
\newblock {\em Journal of biomechanics}, 28(10):1167--1177, 1995.
\newblock DOI: 10.1016/0021-9290(94)00174-3.

\bibitem{aguado2011patient}
Jazmin Aguado-Sierra, Adarsh Krishnamurthy, Christopher Villongco, Joyce
  Chuang, Elliot Howard, Matthew~J Gonzales, Jeff Omens, David~E Krummen,
  Sanjiv Narayan, Roy~CP Kerckhoffs, et~al.
\newblock Patient-specific modeling of dyssynchronous heart failure: a case
  study.
\newblock {\em Progress in biophysics and molecular biology}, 107(1):147--155,
  2011.
\newblock DOI: 10.1016/j.pbiomolbio.2011.06.014.

\bibitem{sermesant2012patient}
Maxime Sermesant, Radomir Chabiniok, Phani Chinchapatnam, Tommaso Mansi,
  Florence Billet, Philippe Moireau, Jean-Marc Peyrat, K~Wong, Jatin Relan,
  Kawal Rhode, et~al.
\newblock Patient-specific electromechanical models of the heart for the
  prediction of pacing acute effects in crt: a preliminary clinical validation.
\newblock {\em Medical image analysis}, 16(1):201--215, 2012.
\newblock DOI: 10.1016/j.media.2011.07.003.

\bibitem{crozier2016relative}
Andrew Crozier, Bojan Blazevic, Pablo Lamata, Gernot Plank, Matthew Ginks,
  Simon Duckett, Manav Sohal, Anoop Shetty, Christopher~A Rinaldi, Reza Razavi,
  et~al.
\newblock The relative role of patient physiology and device optimisation in
  cardiac resynchronisation therapy: a computational modelling study.
\newblock {\em Journal of molecular and cellular cardiology}, 96:93--100, 2016.
\newblock DOI: 10.1016/j.yjmcc.2015.10.026.

\bibitem{kayvanpour2015towards}
Elham Kayvanpour, Tommaso Mansi, Farbod Sedaghat-Hamedani, Ali Amr, Dominik
  Neumann, Bogdan Georgescu, Philipp Seegerer, Ali Kamen, Jan Haas, Karen~S
  Frese, et~al.
\newblock Towards personalized cardiology: multi-scale modeling of the failing
  heart.
\newblock {\em PLoS One}, 10(7):e0134869, 2015.
\newblock DOI: 10.1371/journal.pone.0134869.

\bibitem{hodgkin1952quantitative}
Alan~L Hodgkin and Andrew~F Huxley.
\newblock A quantitative description of membrane current and its application to
  conduction and excitation in nerve.
\newblock {\em The Journal of physiology}, 117(4):500, 1952.
\newblock DOI: 10.1113/jphysiol.1952.sp004764.

\bibitem{nash2000computational}
Martyn~P Nash and Peter~J Hunter.
\newblock Computational mechanics of the heart.
\newblock {\em Journal of elasticity and the physical science of solids},
  61:113--141, 2000.
\newblock DOI: 10.1023/A:1011084330767.

\bibitem{o2011simulation}
Thomas O'Hara, L{\'a}szl{\'o} Vir{\'a}g, Andr{\'a}s Varr{\'o}, and Yoram Rudy.
\newblock Simulation of the undiseased human cardiac ventricular action
  potential: model formulation and experimental validation.
\newblock {\em PLoS computational biology}, 7(5):e1002061, 2011.
\newblock DOI: 10.1371/journal.pcbi.1002061.

\bibitem{ten2008modelling}
K.~H. W.~J Ten~Tusscher and Alexander~V Panfilov.
\newblock Modelling of the ventricular conduction system.
\newblock {\em Progress in biophysics and molecular biology}, 96(1-3):152--170,
  2008.
\newblock DOI: 10.1016/j.pbiomolbio.2007.07.026.

\bibitem{keener2009mathematical}
James Keener and James Sneyd.
\newblock {\em Mathematical physiology: II: Systems physiology}.
\newblock Springer, 2009.
\newblock DOI: 10.1007/978-0-387-79388-7.

\bibitem{bueno2008minimal}
Alfonso Bueno-Orovio, Elizabeth~M Cherry, and Flavio~H Fenton.
\newblock Minimal model for human ventricular action potentials in tissue.
\newblock {\em Journal of theoretical biology}, 253(3):544--560, 2008.
\newblock DOI: 10.1016/j.jtbi.2008.03.029.

\bibitem{iyer2004computational}
Vivek Iyer, Reza Mazhari, and Raimond~L Winslow.
\newblock A computational model of the human left-ventricular epicardial
  myocyte.
\newblock {\em Biophysical journal}, 87(3):1507--1525, 2004.
\newblock DOI: 10.1529/biophysj.104.043299.

\bibitem{lancaster2016improved}
M~Cummins Lancaster and EA~Sobie.
\newblock Improved prediction of drug-induced torsades de pointes through
  simulations of dynamics and machine learning algorithms.
\newblock {\em Clinical Pharmacology \& Therapeutics}, 100(4):371--379, 2016.
\newblock DOI: 10.1002/cpt.367.

\bibitem{mayourian2017experimental}
Joshua Mayourian, Timothy~J Cashman, Delaine~K Ceholski, Bryce~V Johnson, David
  Sachs, Deepak~A Kaji, Susmita Sahoo, Joshua~M Hare, Roger~J Hajjar, Eric~A
  Sobie, et~al.
\newblock Experimental and computational insight into human mesenchymal stem
  cell paracrine signaling and heterocellular coupling effects on cardiac
  contractility and arrhythmogenicity.
\newblock {\em Circulation research}, 121(4):411--423, 2017.
\newblock DOI: 10.1161/CIRCRESAHA.117.310796.

\bibitem{doerschuk1990modelling}
Peter~C Doerschuk, Robert~R Tenney, and Alan~S Willsky.
\newblock Modelling electrocardiograms using interacting markov chains.
\newblock {\em International journal of systems science}, 21(2):257--283, 1990.
\newblock DOI: 10.1080/00207729008910361.

\bibitem{smith2004development}
NP~Smith and EJ~Crampin.
\newblock Development of models of active ion transport for whole-cell
  modelling: cardiac sodium--potassium pump as a case study.
\newblock {\em Progress in biophysics and molecular biology}, 85(2-3):387--405,
  2004.
\newblock DOI: 10.1016/j.pbiomolbio.2004.01.010.

\bibitem{lumens2015differentiating}
Joost Lumens, Bhupendar Tayal, John Walmsley, Antonia Delgado-Montero, Peter~R
  Huntjens, David Schwartzman, Andrew~D Althouse, Tammo Delhaas, Frits~W
  Prinzen, and John Gorcsan~III.
\newblock Differentiating electromechanical from non--electrical substrates of
  mechanical discoordination to identify responders to cardiac
  resynchronization therapy.
\newblock {\em Circulation: Cardiovascular Imaging}, 8(9):e003744, 2015.
\newblock DOI: 10.1161/CIRCIMAGING.115.003744.

\bibitem{ouyang2020video}
David Ouyang, Bryan He, Amirata Ghorbani, Neal Yuan, Joseph Ebinger, Curtis~P
  Langlotz, Paul~A Heidenreich, Robert~A Harrington, David~H Liang, Euan~A
  Ashley, et~al.
\newblock Video-based ai for beat-to-beat assessment of cardiac function.
\newblock {\em Nature}, 580(7802):252--256, 2020.
\newblock DOI: 10.1038/s41586-020-2145-8.

\bibitem{jiwani2021novel}
Nasmin Jiwani, Ketan Gupta, and Pawan Whig.
\newblock Novel healthcare framework for cardiac arrest with the application of
  ai using ann.
\newblock In {\em 2021 5th international conference on information systems and
  computer networks (ISCON)}, pages 1--5. IEEE, 2021.
\newblock DOI: 10.1109/ISCON52037.2021.9702493.

\bibitem{zhang2020real}
Pei-I Zhang, Chien-Chin Hsu, Yuan Kao, Chia-Jung Chen, Ya-Wei Kuo, Shu-Lien
  Hsu, Tzu-Lan Liu, Hung-Jung Lin, Jhi-Joung Wang, Chung-Feng Liu, et~al.
\newblock Real-time ai prediction for major adverse cardiac events in emergency
  department patients with chest pain.
\newblock {\em Scandinavian Journal of Trauma, Resuscitation and Emergency
  Medicine}, 28(1):1--7, 2020.
\newblock DOI: 10.1186/s13049-020-00786-x.

\bibitem{van1928lxxii}
Balthasar Van Der~Pol and Jan Van Der~Mark.
\newblock Lxxii. the heartbeat considered as a relaxation oscillation, and an
  electrical model of the heart.
\newblock {\em The London, Edinburgh, and Dublin Philosophical Magazine and
  Journal of Science}, 6(38):763--775, 1928.
\newblock DOI: 10.1080/14786441108564652.

\bibitem{ryzhii2014heterogeneous}
Elena Ryzhii and Maxim Ryzhii.
\newblock A heterogeneous coupled oscillator model for simulation of ecg
  signals.
\newblock {\em Computer methods and programs in biomedicine}, 117(1):40--49,
  2014.
\newblock DOI: 10.1016/j.cmpb.2014.04.009.

\bibitem{fitzhugh1961impulses}
Richard FitzHugh.
\newblock Impulses and physiological states in theoretical models of nerve
  membrane.
\newblock {\em Biophysical journal}, 1(6):445--466, 1961.
\newblock DOI: 10.1016/S0006-3495(61)86902-6.

\bibitem{nagumo1962active}
Jinichi Nagumo, Suguru Arimoto, and Shuji Yoshizawa.
\newblock An active pulse transmission line simulating nerve axon.
\newblock {\em Proceedings of the IRE}, 50(10):2061--2070, 1962.
\newblock DOI: 10.1109/JRPROC.1962.288235.

\bibitem{di1998model}
Diego di~Bernardo and Maria~G Signorini.
\newblock A model of two nonlinear coupled oscillators for the study of
  heartbeat dynamics.
\newblock {\em International journal of Bifurcation and Chaos},
  8(10):1975--1985, 1998.
\newblock DOI: 10.1142/S0218127498001637.

\bibitem{sato1994bonhoeffer}
S~Sato, S~Doi, and T~Nomura.
\newblock Bonhoeffer-van der pol oscillator model of the sino-atrial node: a
  possible mechanism of heart rate regulation.
\newblock {\em Methods of information in medicine}, 33(01):116--119, 1994.
\newblock DOI: 10.1055/s-0038-1634966.

\bibitem{grudzinski2004modeling}
Krzysztof Grudzi{\'n}ski and Jan~J {\.Z}ebrowski.
\newblock Modeling cardiac pacemakers with relaxation oscillators.
\newblock {\em Physica A: statistical Mechanics and its Applications},
  336(1-2):153--162, 2004.
\newblock DOI: 10.1016/j.physa.2004.01.020.

\bibitem{gois2009analysis}
Sandra~RFSM Gois and Marcelo~A Savi.
\newblock An analysis of heart rhythm dynamics using a three-coupled oscillator
  model.
\newblock {\em Chaos, Solitons \& Fractals}, 41(5):2553--2565, 2009.
\newblock DOI: 10.1016/j.chaos.2008.09.040.

\bibitem{purves2019neurosciences}
Dale Purves, George~J Augustine, David Fitzpatrick, William Hall,
  Anthony-Samuel LaMantia, and Leonard White.
\newblock {\em Neurosciences}.
\newblock De Boeck Sup{\'e}rieur, 2019.
\newblock ISBN: 0-87893-742-0.

\bibitem{litfl}
Mike Cadogan and Robert Buttner.
\newblock Ecg library.
\newblock \url{https://litfl.com/ecg-library/}.

\bibitem{kusumoto2020ecg}
Fred Kusumoto.
\newblock {\em ECG interpretation: from pathophysiology to clinical
  application}.
\newblock Springer Nature, 2020.
\newblock DOI: 10.1007/978-0-387-88880-4.

\bibitem{wagner2001marriott}
Galen~S Wagner.
\newblock {\em Marriott's practical electrocardiography}.
\newblock Lippincott Williams \& Wilkins, 2001.
\newblock ISBN: 978-1451146257.

\bibitem{santana2010does}
Luis~F Santana, Edward~P Cheng, and W~Jonathan Lederer.
\newblock How does the shape of the cardiac action potential control calcium
  signaling and contraction in the heart?
\newblock {\em Journal of molecular and cellular cardiology}, 49(6):901, 2010.
\newblock DOI: 10.1016/j.yjmcc.2010.09.005.

\bibitem{hampton2019ecg}
John Hampton and Joanna Hampton.
\newblock {\em The ECG made easy e-book}.
\newblock Elsevier Health Sciences, 2019.
\newblock ISBN: 978-0-7020-7457-8.

\bibitem{morris1981voltage}
Catherine Morris and Harold Lecar.
\newblock Voltage oscillations in the barnacle giant muscle fiber.
\newblock {\em Biophysical journal}, 35(1):193--213, 1981.
\newblock DOI: 10.1016/S0006-3495(81)84782-0.

\bibitem{west1985nonlinear}
Bruce~J West, Ary~L Goldberger, Galina Rovner, and Valmik Bhargava.
\newblock Nonlinear dynamics of the heartbeat: I. the av junction: Passive
  conduit or active oscillator?
\newblock {\em Physica D: Nonlinear Phenomena}, 17(2):198--206, 1985.
\newblock DOI: 10.1016/0167-2789(85)90004-1.

\bibitem{PhysioNet}
A.~L. Goldberger, L.~A.~N. Amaral, L.~Glass, J.~M. Hausdorff, P.~Ch. Ivanov,
  R.~G. Mark, J.~E. Mietus, G.~B. Moody, C.-K. Peng, and H.~E. Stanley.
\newblock {PhysioBank, PhysioToolkit, and PhysioNet}: Components of a new
  research resource for complex physiologic signals.
\newblock {\em Circulation}, 101(23):e215--e220, 2000 (June 13).
\newblock Circulation Electronic Pages:
  http://circ.ahajournals.org/content/101/23/e215.full PMID:1085218; DOI:
  10.1161/01.CIR.101.23.e215.

\end{thebibliography}
\end{document}